# Stylolites: a review

Toussaint R.[1,2,3*], Aharonov E.[4], Koehn, D.[5], Gratier, J.-P.[6], Ebner, M.[7], Baud, P.[1], Rolland, A.[1], and Renard, F.[6,8]

[1]*Institut de Physique du Globe de Strasbourg, CNRS, University of Strasbourg, 5 rue Descartes, F-67084 Strasbourg Cedex, France. Phone : +33 673142994. email : renaud.toussaint@unistra.fr*

[2] *International Associate Laboratory D-FFRACT, Deformation, Flow and Fracture of Disordered Materials, France-Norway.*

[3]*SFF PoreLab, The Njord Centre, Department of Physics, University of Oslo, Norway.*

[4]*Institute of Earth Sciences, The Hebrew University, Jerusalem, 91904, Israel*

[5]*School of Geographical and Earth Sciences, University of Glasgow, UK*

[6]*University Grenoble Alpes, ISTerre, Univ. Savoie Mont Blanc, CNRS, IRD, IFSTTAR, 38000 Grenoble, France*

[7]*OMV Exploration & Production GmbH Trabrennstrasse 6-8, 1020 Vienna, Austria*

[8] *The Njord Centre, PGP, Department of Geosciences, University of Oslo, Norway*

*\*corresponding author*

**Highlights:**

. Stylolite formation depends on rock composition and structure, stress and fluids.

. Stylolite geometry, fractal and self-affine properties, network structure, are investigated.

. The experiments and physics-based numerical models for their formation are reviewed.

. Stylolites can be used as markers of strain, paleostress orientation and magnitude.

. Stylolites impact transport properties, as function of maturity and flow direction.

## Abstract

Stylolites are ubiquitous geo-patterns observed in rocks in the upper crust, from geological reservoirs in sedimentary rocks to deformation zones, in folds, faults, and shear zones. These rough surfaces play a major role in the dissolution of rocks around stressed contacts, the transport of dissolved material and the precipitation in surrounding pores. Consequently, they play an active role in the evolution of rock microstructures and rheological properties in the Earth's crust. They are observed individually or in networks, in proximity to fractures and joints, and in numerous geological settings. This review article deals with their geometrical and compositional characteristics and the factors leading to their genesis. The main questions this review focuses on are the following: How do they form? How can they be used to measure strain and formation stress? How do they control fluid flow in the upper crust?

Geometrically, stylolites have fractal roughness, with fractal geometrical properties exhibiting typically three scaling regimes: a self-affine scaling with Hurst exponent 1.1+/-0.1 at small scale (up to tens or hundreds of microns), another one with Hurst exponent around 0.5 to 0.6 at intermediate scale (up to millimeters or centimeters), and in the case of sedimentary stylolites, a flat scaling at large scale. More complicated anisotropic scaling (scaling laws depending of the direction of the profile considered) is found in the case of tectonic stylolites. We report models based on first principles from physical chemistry and statistical physics, including a mechanical component for the free-energy associated with stress concentrations, and a precise tracking of the influence of grain-scale heterogeneities and disorder on the resulting (micro)structures. Experimental efforts to reproduce stylolites in the laboratory are also reviewed. We show that although micrometer-size stylolite teeth are obtained in laboratory experiments, teeth deforming numerous grains have not yet been obtained experimentally, which is understandable given the very long formation time of such geometries. Finally, the applications of stylolites as strain and stress markers, to determine paleostress magnitude are reviewed. We show that the scalings in stylolite heights and the crossover scale between these scalings can be used to determine the stress magnitude (its scalar value) perpendicular to the stylolite surface during the stylolite formation, and that the stress anisotropy in the stylolite plane can be determined for the case of tectonic stylolites. We also show that the crossover between medium (millimetric) scales and large (pluricentimetric) scales, in the case of sedimentary stylolites, provides a good marker for the total amount of dissolution, which is still valid even when the largest teeth start to dissolve – which leads to the loss of information, since the total deformation is not anymore recorded in a single marker structure. We discuss the impact of the stylolites on the evolution of the transport properties of the hosting rock, and show that they promote a permeability increase parallel to the stylolites, whereas their effect on the permeability transverse to the stylolite can be negligible, or may reduce the permeability, depending on the development of the stylolite.

# 1. Introduction

Stylolites are natural rock-rock interlocked interfaces that may produce spectacular rough patterns observed in field outcrops, cores and in stones used for buildings (Figure 1). They form by a localized dissolution process and their interface contains minerals at concentrations different from that in the surrounding host rock. The term *stylolite* – a combination of the ancient Greek words *stylos* (pillar) and *lithos* (stone) – was coined in the 19$^{th}$ century to refer to the observation that these patterns may display columnar shapes (Klöden, 1828, cited in Shaub, 1939). A classification of the various stylolite patterns and networks was proposed by Park and Schot (1968). In this introduction, we present how the ideas evolved from the 19$^{th}$ century to the middle of the 20$^{th}$ century about the observations of stylolites and their genesis. The developments over the past fifty years are then discussed in the different sections of this review article.

Stylolites were originally attributed to organic species (Klöden, 1828) and later described as related to diagenetic processes in various sedimentary rocks: mainly carbonates (Hopkins, 1897; Gordon, 1918) and quartz-cemented quartzarenite (called quartzite by Tarr (1916)). Sorby (1862) was the first to attribute them to pressure-solution, by describing them as "curious teeth-like projections with which one bed of limestone enters into another". The interest in stylolites increased at the beginning of the 20$^{th}$ century with the work of Stockdale (1922, 1926), who defended his master thesis on stylolites in Indiana limestones in 1921. He recognized that the insoluble minerals found in the stylolites were solution residues from the surrounding limestone and were not related to a temporary change in the conditions of sedimentation. He also observed that distortions or displacements of the rock around the stylolite were absent, which ruled out pure mechanical penetration, leading him to propose a complete theory by which these rough surfaces formed by dissolution. He was also the first to suggest that the length of a tooth in a stylolite would represent only a minimum thickness of dissolution (Figure 1g) and proposed that stylolites form when the rock was already consolidated (Stockdale, 1926). He concluded that, because the displacement due to dissolution occurs in a slow movement, the stylolitic columns are striated, showing slickensides. His interpretation was also that water charged with carbon dioxide could circulate along the stylolite to enhance the dissolution of the limestone, which was called the *solution model*, where the interface was pre-existing (e.g. sedimentary bedding). Later, the *pressure solution model* was proposed (Weyl, 1959; Heald, 1959; Trurnit, 1968), where dissolution is related to the coupling between chemical and mechanical forces, as proposed by Sorby (1862), based on the theory of non-equilibrium thermodynamics of Gibbs (1877).

The solution model was questioned by Shaub (1939), among others, who discussed an alternative explanation where dissolution did not play a role and proposed that stylolites formed only in soft sediments during compaction, under the effect of fluid expulsion and plastic flow. Although the scientific debate was vigorous (e.g., Dunnington, 1954), it is now widely accepted that stylolites form within solid rocks, during diagenesis and/or tectonic deformation. Because of local dissolution along the stylolite, the removed material may provide the cement that precipitates in the host rock surrounding the stylolites, as proposed by Heald (1956, 1959) for sandstones. Moreover, Heald (1959) also proposed that clays may enhance the dissolution along the stylolite and that some clay minerals are necessary for stylolites to develop. Bjorkum (1996) also observed that the presence of illitic or micaceous clay is necessary for the dissolution of quartz in sandstones, usually described as pressure-solution, and proposed to call this process clay-induced dissolution.

The presence of stylolites, with various amounts of clays along the interface, may control fluid flow in sedimentary formations. For example, Dunnington (1954) mentioned that if stylolites are related to compaction, then fluids must be expelled from the rock. He also noted that, because of compaction, stylolites could act as seals and stop the upward migration of hydrocarbons. Park and Schot (1968) noted the close relationships between stylolites and minerals deposited by a percolating fluid. Stylolites were thus suggested to impede flow in the host rock, both due to the clay parting that lines them, and due to the reduced permeability that surrounds them (e.g., Heald, 1959; Burgess & Peter, 1985; Koepnick, 1987). The idea of a reduced permeability due to stylolites (Heald, 1959) remained in the literature until it was shown that they may act as conduits by later fieldwork (Lind, 1993; Carozzi and von Bergen, 1987; Korneva et al., 2014; Rustichelli et al., 2015) and laboratory measurements (Heap et al., 2014). Recent observations and numerical models indicate that the stylolite permeability is anisotropic and that stylolites may act both as seals and fluid pathways depending on the material that collects in the stylolite and the offset of sealing material at teeth (Koehn et al., 2016).

In 1922, Stockdale wrote in the introduction of his master thesis that "There are few of the minor, yet important, geologic phenomena whose explanation has been as unsatisfactory and under as much controversy as that of stylolites". More than ninety years later, we propose here an updated review on stylolites to answer the following questions: How do they form? How can they be used to measure strain and stress? How do they control fluid flow in the upper crust? The present review article is organized in sections that present scientific questions related to stylolites using the context of mechanisms of pattern formation, strain indicators and piezometers, and effect on rock permeability. Our conclusion is that, if most scientific controversies are now closed since 1922, some still remain and several directions of further research can be proposed.

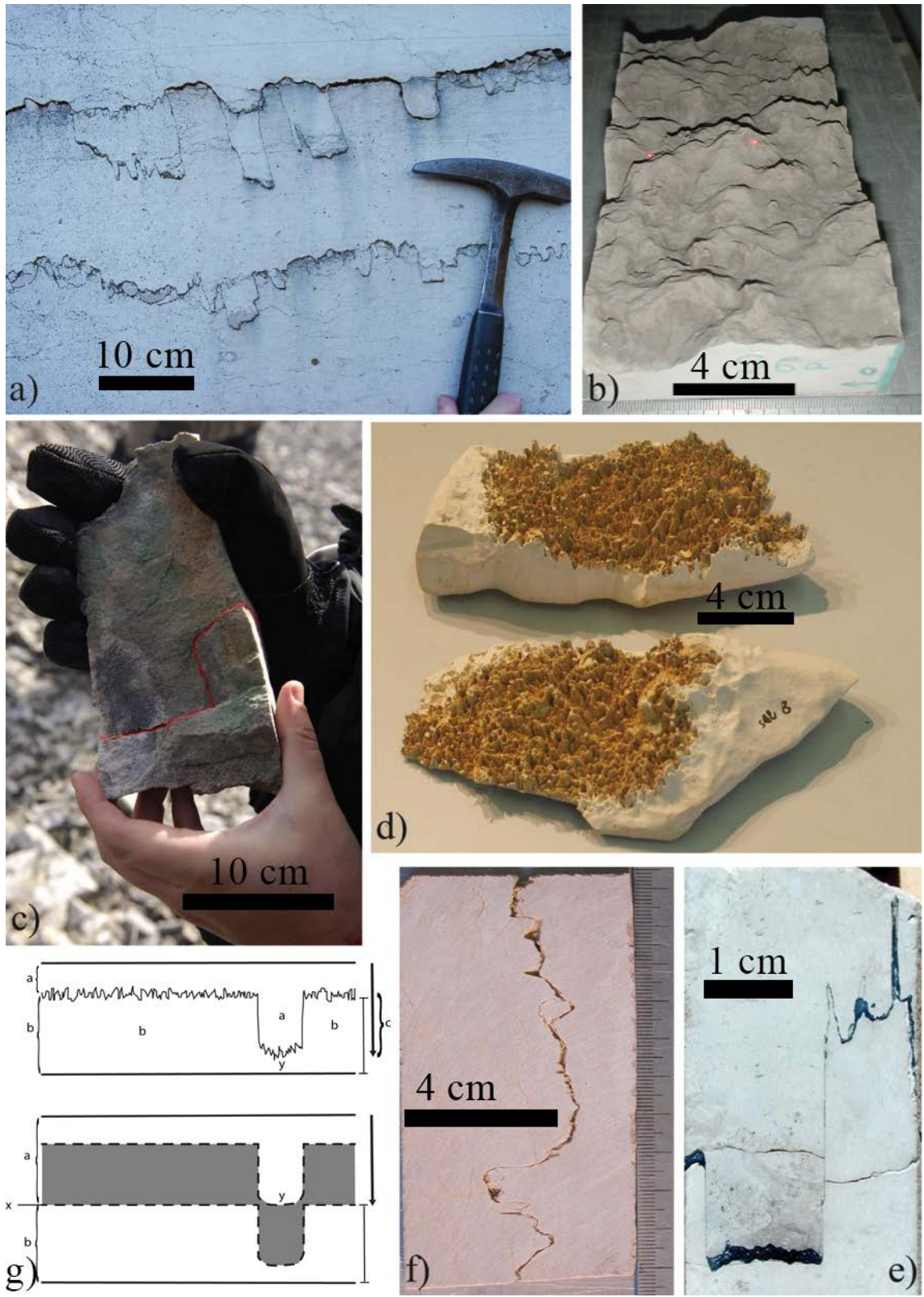

Figure 1: Variety of stylolite patterns. a-b) Columnar and rough stylolites in limestones (Burgundy, France). c) Columnar stylolite in a dolostone (Murchinson bay, Spitzbergen). The stylolite is underlined in red. d) Three-dimensional view of a stylolite in a limestone (Vercors,

France). e) Columnar stylolite with concentration of clays on the top and bottom of teeth (Greece) f) Columnar stylolite in limestone where most teeth have been dissolved (Vercors, France). g) First interpretation that stylolites form due to dissolution along an interface, and that the length of the columns represents a minimum for the thickness of dissolved rock (modified from Stockdale, 1926).

## 2. What are stylolites? Field observations and laboratory characterization

*2.1 Appearance and morphology of stylolites*

Stylolites appear on outcrops as rough dark lines, in numerous sedimentary rocks and in deformed zones of folds, faults, and shear zones. They are found notably, in carbonates such as limestones and marbles (Stockdale, 1922, 1926, 1936, 1943; Dunnington, 1954; Bushinskiy, 1961; Park and Schot, 1968; Bathurst, 1971; Buxton and Sibley, 1981; Railsback, 1993; Safaricz 2002; Safaricz and Davidson 2005; André, 2010; Vandegiste & John, 2013; Rolland et al., 2012, 2014,), cherts (Bushinskiy, 1961; Iijima, 1979; Cox and Whitford-Stark, 1987), coal (Stutzer, 1940), and sandstones (Tarr 1916; Young, 1945; Heald, 1955, 1956; Sibley and Blatt 1976; Stone & Siever, 1996). Pressure solution and cleavage is also observed in shales (Wright and Platt, 1982; Rutter, 1983). The dark aspect is due to the accumulation of non-soluble (often clay-rich) residuals.

Stylolites can be divided into three subcategories: sedimentary stylolites, tectonic stylolites, and slickolites. The average plane of sedimentary stylolites is sub-parallel to bedding – which is horizontal at the time of formation. Surfaces of tectonic stylolites are usually perpendicular to the largest compressive principal stress axis, which can be horizontal, leading to vertical stylolites. They sometimes intersect previously formed sedimentary stylolites. Slickolites develop on planes that are oblique to the largest principal stress direction. Stylolites show large teeth-like structures (Figs. 2, 3), a few mm to cm in size, pointing in the direction of largest compressive stress during their formation (Koehn et al., 2007). The size of the teeth exceeds the size of grains in the rock (Bathurst, 1987), which allows identification of displacement of the pre-existing structure along the sides of the teeth, showing stylolites to bound regions of undissolved material and replace the dissolved one (Rolland et al., 2014), (see Fig. 2 and Supplementary Video for optical and X-ray micro-tomography microscopic views and Fig. 3 for a scanning electron microscopy views). The layer of insoluble elements may also contain some porosity (Fig. 2g-h).

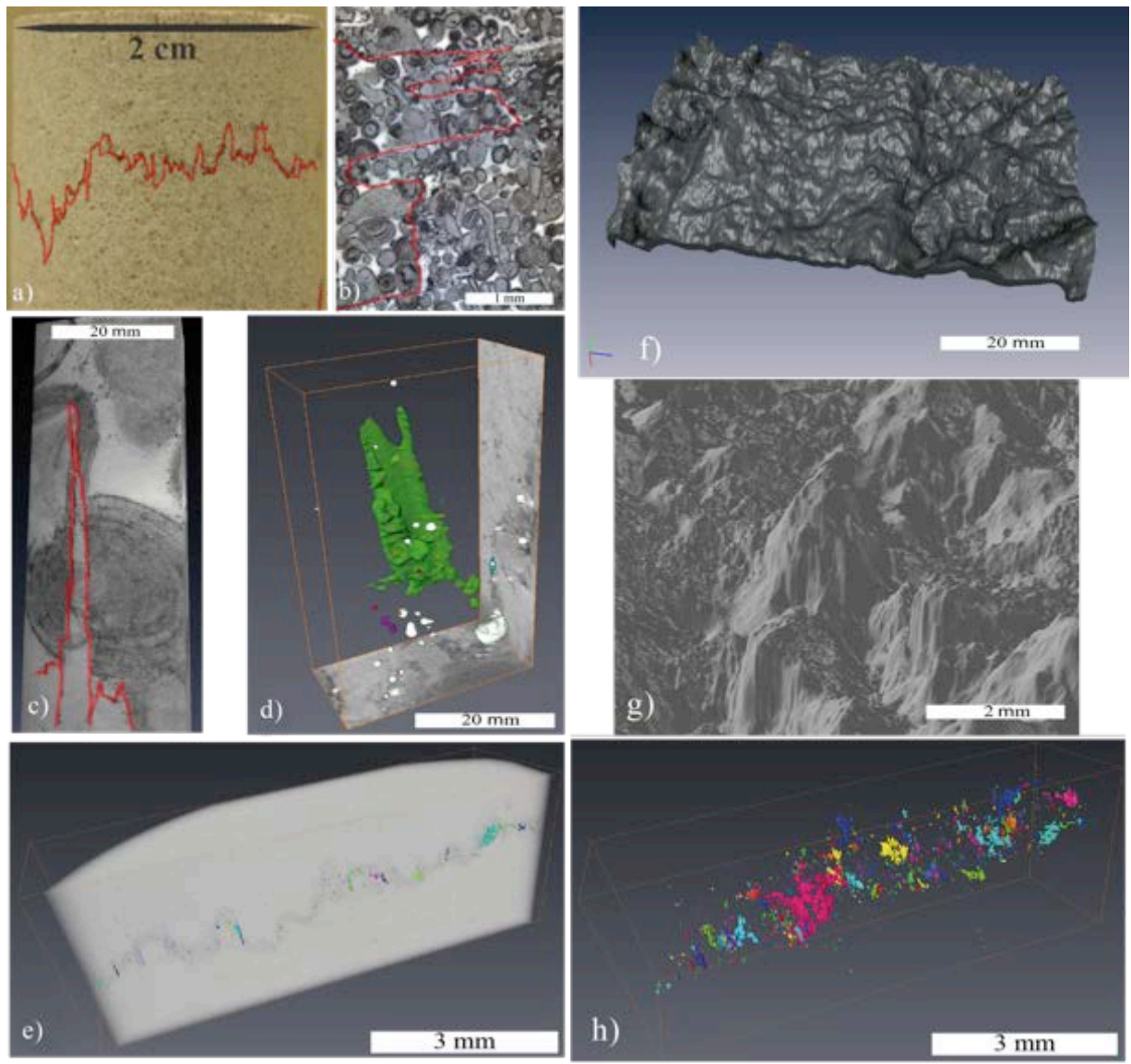

Figure 2: a-b) Stylolite in an oolitic limestone, after Rolland (2013). The red line underlines the stylolites in a,b,c. a) Hand sample from a core log, 2 cm sample size. b) Optical microscopic observation. The stylolite has removed part of the oolites by dissolution. The remaining part of the oolites are sheared by the teeth of the stylolite. c-d) High resolution view (0.7 micron voxel size, beamline ID19, ESRF) of stylolite teeth. In c), a stylolite tooth penetrates in an oolite displaying concentric rings of high density. e and h) 3D view of the stylolite insoluble elements and the porosity (6.27 micron voxel size, beamline ID19, ESRF). Individual pores determined by connectivity properties are underlined by the use of arbitrary different colors for each of them. The rocks in a-h are Oxfordian carbonate formations from Bure-sur-Meuse, France f) 3D tomography of a stylolite in calcite (Greece). The dark part corresponds to the insoluble material. g) Zoom-in on the stylolite shown in f). See also supplementary video for a 3D laboratory tomograph view of the insoluble elements layer in anastomosed stylolites, in a 2cm radius and 4 cm long cylindrical sample from Dogger carbonate formations from Bure-sur-Meuse, France. Dark color corresponds to the denser insoluble material.

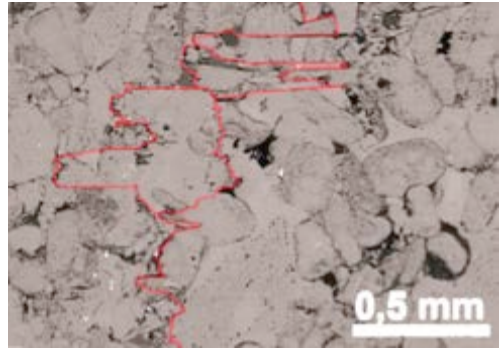

Figure 3: Scanning electron microscopy view of a thin section of Dogger limestone in the Paris Basin, extracted from a borehole of the Bure Underground Research Laboratory. Sample with stylolites, marked in red, where white spots correspond to pyrite and porosity appears in dark.

These average planes of dissolution seams can be oriented parallel to bedding for sedimentary stylolites (Ebner 2009, see also Fig. 4), or vertically or in any other direction for tectonic stylolites (Ebner et al., 2010b, and Fig. 5).

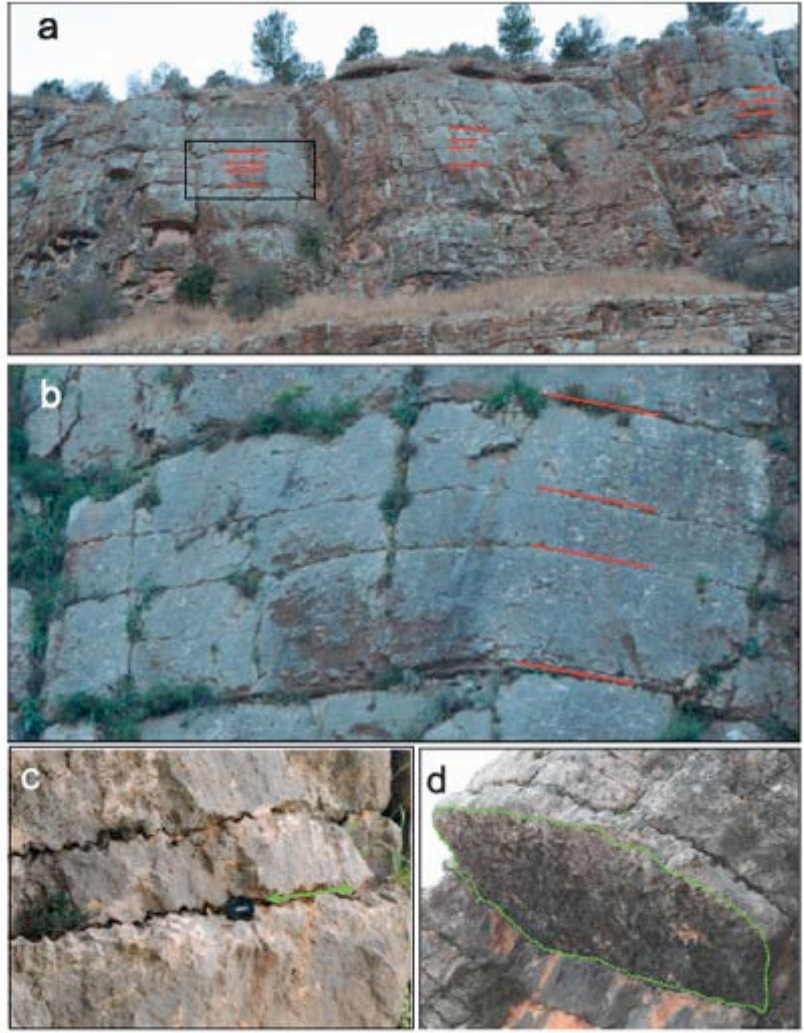

Figure 4: 1 km long parallel stylolites in Blanche cliff, Northern Israel. (a) General view of the cliff with four stylolites (marked in red) showing they are nearly parallel, follow bedding planes and can be traced for a large distance (height of cliffs 50 m, rectangle vertical size 10 m). (b) Zoom on rectangle marked in (a). (c) Zoom-in on two other stylolites, where stylolite cm-scale roughness is evident. (d) An exposed rough stylolite surface. Length of exposed area: 2.3 m. After Ben-Itzhak et al (2012).

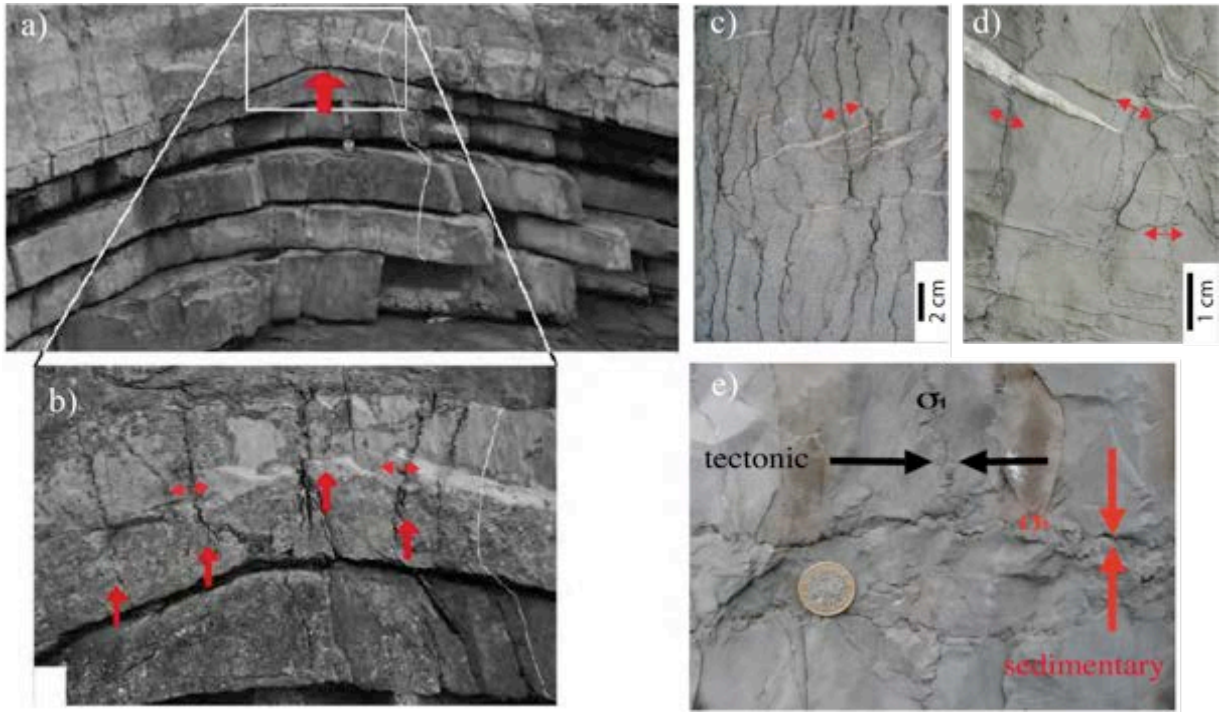

Figure 5: Tectonic stylolites. a), b) tectonic stylolites in different rocks, pointed to by single arrows. Sub-horizontal tensile joints with recrystallized material can be seen. Vertical size of (a): 80 cm, (b): 12 cm. c): Field view of coexisting tectonic and sedimentary stylolites, formed by successive stress configurations, where horizontal stress direction becoming largest after tectonic loading, following an initial sedimentary configuration where the largest stress direction was vertical. Double arrows (pointing indifferently in or out) indicate the orientation of the largest principal stress at the time of formation of the corresponding stylolites.

They can also be found to accommodate local stress perturbations at small scale, such as in fault breccias (Géraud, 2006) or fault gouges (Chambon et al., 2006; Gratier et al., 2011).

In porous rocks, other structures of localized deformation also perpendicular to the principal stress axis can form, such as pure compaction bands (Mollema and Antonellini, 1996; Olsson, 1999; Baud et al., 2004; Fossen et al., 2011; Cheung et al., 2012). These bands differ from stylolites, forming brittlely with grain crushing. Compaction bands can also develop during shear, where they are termed shear-enhanced compaction bands and are oblique to the largest compressive stress axis. In such a case, they can develop with stylolites perpendicular to this direction (see e.g. Fossen et al., 2011, Fig. 3 in this reference).

When a stylolite grows along a pre-existing interface that is oblique to $\sigma_1$, the largest principal stress axis, inclined teeth form that are not perpendicular to their average plane, also termed slickolites (Stockdale, 1922; Ebner, 2010b). Since the main stress axis is not perpendicular to the surface, the displacement between the two undissolved rock blocks bounding the slickolite involves components of surface-normal shortening and surface-parallel sliding (Fig. 6).

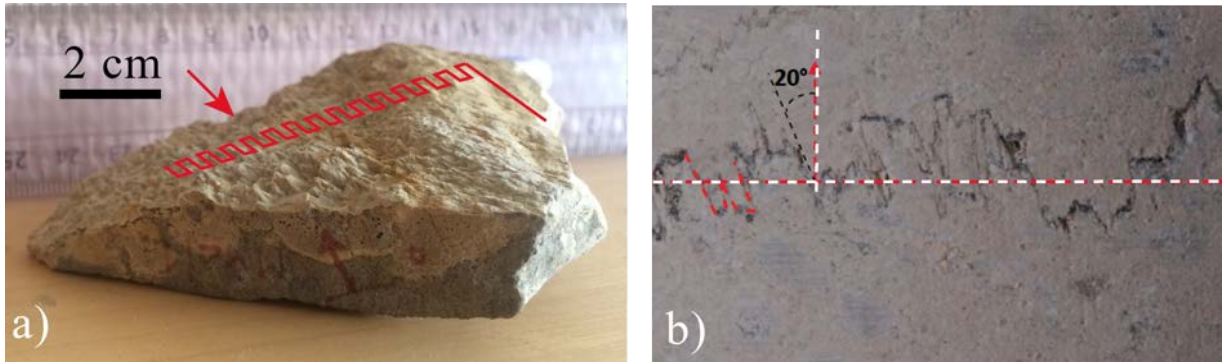

Fig. 6: a) Open slickolite. The main orientation of the teeth (arrows and red lines) is not perpendicular to the average plane. b) Slickolite on a core log (3cm long view): the teeth are inclined by about 20° from the normal to the mean plane. This geometry indicates that the main stress direction during the formation of these teeth was inclined either because it occurs on a preexisting surface oblique to maximum compressive stress (see Fig. 22) or because of a rotation of stress following the initial stylolite formation. Adapted after (Rolland, 2013).

When the thickness of the layer rich in clay and other insoluble elements reaches few micrometers to millimeters, it is sometimes possible to examine the actual surface of the stylolite by mechanically opening the stylolitic surface through the use of a moderate tension perpendicularly to its average direction, by hand or using a handtool. The resulting rough surface is shown on Figs. 1b,d and 2b (Renard et al., 2004; Ebner et al., 2010b; Rolland et al., 2012). This allows measurements of the surface height topography using profilometers or three-dimensional reconstruction from stereophotogrammetry, and spectral analysis of the profiles along different directions along the average plane.

## 2.2 Organization of stylolites in networks and influence of fractures and veins

Sedimentary stylolites can be found isolated, but most of the time they are observed in groups of sub-parallel planar dissolution seams, with bed-parallel lengths of meters to hundreds of meters, while maintaining a nearly constant spacing between them (following bedding planes) as seen in Fig. 4 (Ben-Itzhak et al. 2012), or anastomosing, i.e. merging with each other (Ben-Itzhak et al., 2014) as in Fig. 7a. They are also often found together with groups of fractures or veins (e.g. Milliken, 1994; Peacock and Sanderson, 1995; Smith, 2000; André, 2010; Ben-Itzhak et al, 2014), as seen in Fig. 7b. Stylolites can be found to interact with other stylolites, starting on the teeth corners. Veins can also be found starting on such corners, or lying subparallel to the average stylolite surface and intersecting the sides of the stylolite teeth (Ben-Itzhak et al., 2014; Heap et al., 2014). Indeed, stylolites act as stress concentrators, as can be demonstrated in numerical simulations (Zhou and Aydin, 2010, 2012; Katsman, 2010; Rolland et al., 2012), or as can be seen by the geometrical relations between fissures and stylolites seen in outcrops (Fig. 7b and Zhou and Aydin, 2012), grain scale imaging by cathodoluminescence (CL) (Milliken, 1994; Dickinson and Milliken, 1995; Land and Milliken, 2000; Makowitz and Milliken 2003), or from fractures formed during laboratory uniaxial-deformation tests on limestone samples containing stylolites (Fig. 8, see also Rolland 2013). Compressive areas in faults are also well known locations for secondary tectonic stylolite formation (Rispoli, 1981; Fletcher and Pollard 1981; Willemse et al, 1997; Tondi et al, 2006).

Stylolites exhibit several important scaling behaviors: Long (pluricentimetric) parallel stylolites (as in Fig. 4) are fractal, with self-affine roughness properties (e.g. Karcz and Scholz 2003 and references reviewed in sec 2.4). In addition, individual pressure solution seams may display a scaling of length to thickness, as shown by Nenna and Aydin (2012) (Fig. 12 in this reference) for bedding-perpendicular tectonic solution seams.

Beyond the fractal character of individual stylolites, (see below Sec. 2.4), it is possible to also analyze anastamosing networks (Fig. 7a) and their collective organization. It was found (Ben Itzhak 2014; Kaduri, 2013) that the islands of undissolved material between the stylolites also display a scaling law with a power-law size distribution. This feature is similar to regions opened during avalanches between successive positions of pinned fracture fronts (Tallakstad et al. 2011). The aspect ratio between the two principal directions also scales with the island size, which is also similar to fracture avalanche regions (Måløy et al. 2006).

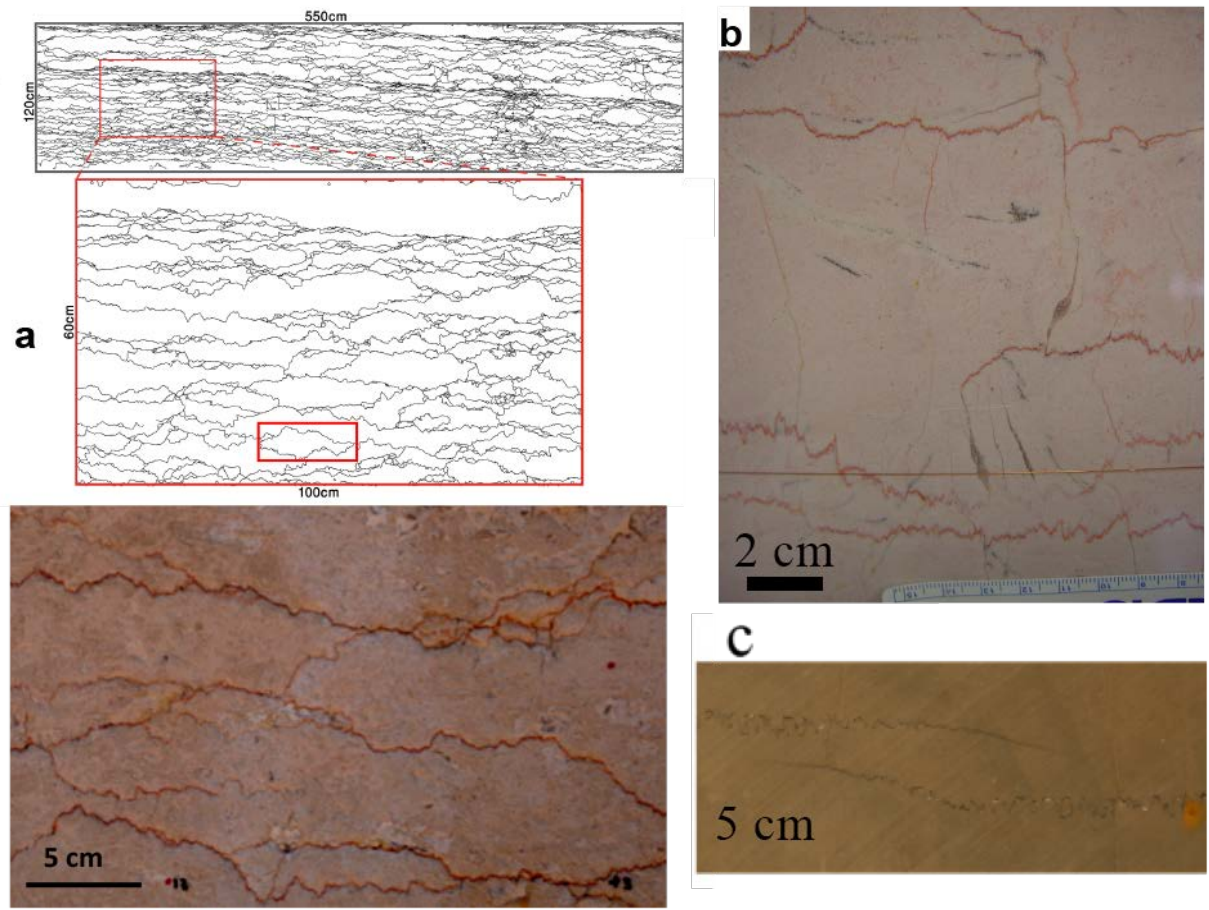

Figure 7: Two types of interconnected networks of stylolites, after (Ben-Itzhak et al., 2014): a) *Anastomosing stylolite network* from a Cenomanian limestone near Mitzpe Ramon, Israel. Three successive zooms reveal a wide (actually self-similar) distribution of "island" sizes, where an island is defined as a region of un-dissolved rock enclosed by stylolitic surfaces (e.g. see island bound by red rectangle in mid panel) b) *Interconnected network of stylolites and veins* in a Jurassic limestone (Calcare Massicio Formation, Italy). The termination of stylolites is typically the beginning of a normal-to-the-stylolite vein. Veins often terminate on stylolites' extremities or on stylolites' teeth corners. c) Two terminations of stylolites interacting with each other, as shows their deflection and reduction of the teeth amplitude towards the terminations.

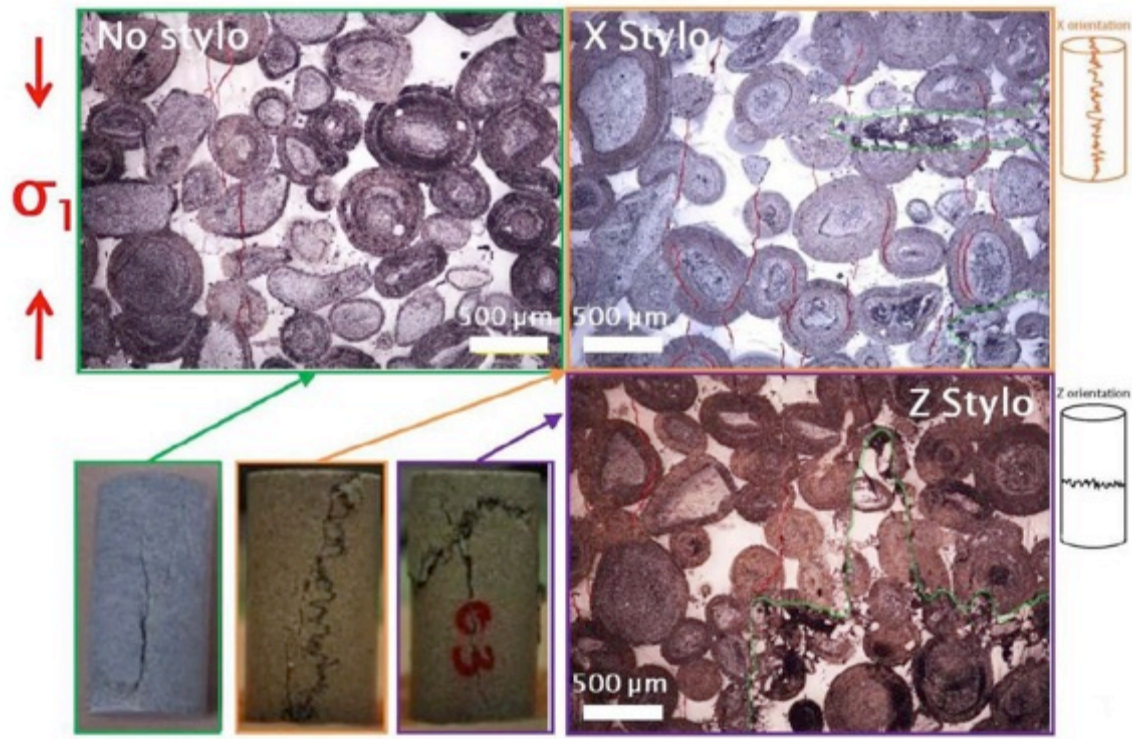

Figure 8: Effect of stylolite on experimental fracture development. Thin sections in Oxfordian formation in the Paris Basin, Bure underground laboratory (after Rolland, 2013), after a uniaxial deformation test carried beyond peak stress: Top left: sample without stylolite. Bottom left: broken 2 cm diameter samples after the tests, corresponding to the thin sections shown in the three other panels. Top right: sample with teeth oriented along the major stress axis. Bottom right: sample with teeth aligned with the major stress axis. Fissures (red) appeared during the test, mostly parallel to the axis of the largest principal compressive stress, $\sigma_1$. They are concentrated around the stylolite and the stylolite teeth corners.

*2.3 Microstructural and chemical analyses: what can be found in the residuals contained in a stylolite?*

The coating or parting along the two opposing rock surfaces that compose a stylolite was recognized by Stockdale in 1922 as a residual of the dissolved rock, which he called a "residual clay seam". Most subsequent authors (e.g. Heald, 1955; Dunnington, 1954; Park and Schot, 1968; Railsback, 1993) followed Stockdale's initial idea, although for some rocks a primary origin of the clay parting has been assumed (Oldershaw and Scoffin, 1967). Stockdale's conclusion is based on the assumption of a homogeneous distribution of insoluble material in the host rock. This material could potentially originate both from initial clay-rich laminae, and/or from a superimposed initial homogeneous distribution. The material is passively enriched in clay during the dissolution process since clay is insoluble, and its clay concentration depends on the host rock composition and can thus vary significantly. Park and Schot (1968) described various stylolite filling materials, such as bituminous material, clay minerals, quartz, dolomite, sulfides (e.g. pyrite) and fluorite for some limestones in the USA. In addition it was found by Railsback (1993) and Andrews and Railsback (1997) that sedimentary carbonate grains form what they call "penetrating particles" along the stylolite peaks. They report that these grains are enriched in stylolite-filling materials along the stylolite compared to the cement phase (Fig. 2). Thus, carbonate grains can also form a part of the stylolite filling in a limestone. Stylolite residual in chalk is similar to that of limestones, comprising clay minerals (illite-smectite), quartz (chert), pyrite, dolomite and apatite (Lind, 1993; Fabricius, 2007; Fabricius and Borre, 2007). For sandstones, stylolite-filling materials are feldspars, micas, clay minerals, oxides, sulfides and other dense minerals (Heald, 1955; Tada and Siever, 1989; Harris, 2006). In general, the main constituent of the residual noted by all of these authors are clay minerals for limestones, and dolomites, and clay and/or mica minerals for sandstones. In the following sections, we will use the term residual for all passively enriched constituents along a stylolite in a genetic sense only, because the exact composition found along such pressure solution interfaces is dependent on host rock composition.

Although most residuals can be assumed to be passively enriched during dissolution, ample proof exists that authigenic mineral can grow in the stylolite residual (Merino et al., 1983; Thomas et al., 1993), and additional minerals may form during secondary fluid flow along the stylolite interface (Braithwaite, 1989), sometime leading to mineable ore concentration (Gratier et al. 2013a). Viti et al. (2014) studied tectonic stylolite parting from four different locations in Italy. The seams were found to be filled with ultrafine clay matrix, enclosing larger grains of relict calcite, insoluble minerals and newly precipitated phases. In all samples, the ultrafine matrix exhibits abundant fissuring and delamination. The seams revealed the occurrence of other nanophases, mostly formed by precipitation from fluids circulating in syn- or post-tectonic regimes, including Fe oxide/hydroxide flakes, $TiO_2$ and apatite nanocrystals. A detailed chemical composition of stylolitic content was also obtained by Evans and Elmore (2006).

It is a common observation that pressure solution features such as stylolites are more common in carbonates than in sandstones (Park and Schot, 1968; Tada and Siever, 1989 and references therein). This relative abundance is consistent with the susceptibility of minerals to pressure solution, first described by Heald (1955), who showed that at temperatures below 250 °C, carbonate minerals are indeed more soluble than quartz (Gratier et al., 2013a), whereas above such temperatures, the solubility of quartz is greater than for calcite. The dissolution series was later refined by Trurnit (1968) who argued that for a sandstone, any mineral less susceptible to dissolution than quartz can be found in the residual (feldspars, oxides, clay and mica minerals, and sulphides). Renard et al. (2004) conducted X-ray fluorescence analysis on stylolite residual and host rock of limestones. They found that stylolites are enriched in aluminum, iron, titanium, and phosphorus compared to the bulk rock, whereas the interface is depleted in calcium indicating preferential dissolution along the stylolite. Concentration ratios for aluminum, iron, titanium, and phosphorus for the residual ranges from 5 – 20 times to that of the host rock.

Several studies (Heald, 1955; Braithwaite, 1986; Tada and Siever, 1989; Railsback, 1993) have shown that usually the most insoluble material can be found at the peaks of stylolites and speculated that these particles could cause the stylolite teeth formation. Clear evidence was only recently presented (Ebner et al., 2010a), confirming that insoluble particles initially present in the host rock may be the cause for stylolite roughening and initiation (Figure 9). Stylolites that have not formed a continuous parting reveal small clay particles at the top of most stylolite peaks. For asperities not occupied by a visual clay particle, a heterogeneity on a smaller scale (i.e. nanometer to atomic) was assumed (Ebner et al., 2010a).

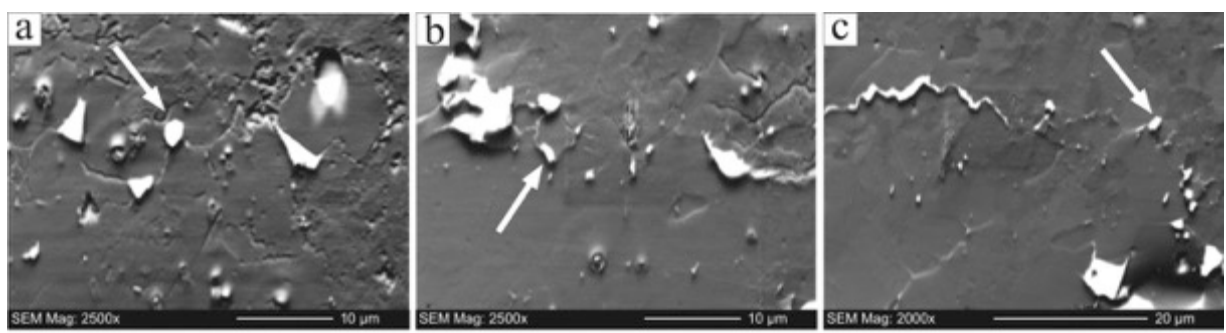

Figure 9**:** Scanning electron microscopy images with orientation contrast of early stage (low strain) sedimentary stylolites in fine-grained Cretaceous limestones from the Cirque de Navacelle, France. Residual is shown to concentrate at stylolite teeth. White spots indicate

clay particles that are found pinned on stylolite's teeth, and the dark gray groundmass is calcite. Stylolites striking roughly E-W in image a, b, c (modified from Ebner et al., 2010a).

Utilizing their observational evidence that clay play a key role in the roughening of stylolites, Ebner et al. (2010a) confirmed an analytical model for stylolite roughening (Renard et al., 2004; Schmittbuhl et al., 2004; Rolland et al., 2012). This model assumes that the cause for the roughening during dissolution is a quenched noise, i.e. compositional or structural heterogeneities, that introduce randomness in physical properties – "noise" – not varying in time – "quenched". This noise is present before the onset of dissolution. The model also shows that the roughening process is balanced by two inherent parameters – surface and elastic energies. This theory was corroborated by numerical models based on a similar approach (Koehn et al., 2007; Ebner et al., 2009a; see also Section 3).

The thickness of the residual was shown to positively correlate with the amount of dissolution (Railsback, 1993; Tada and Siever, 1989). A positive correlation between stylolite amplitude and residual thickness was reported by Heald (1955) and Kaplan (1976), whereas Peacock and Azzam (2006) observed only a very weak correlation between amplitude and residual thickness for a large dataset on dolomite and limestone. The positive relationship between residual thickness and stylolite amplitude seems to be limited to relatively clean rocks, because comparatively thick clay seams have a negative influence on the stylolite amplitude (Park and Schot, 1968). Moreover, clay-rich (Buxton and Sibley, 1981) or mica-rich (Bjorlum, 1996) sections seem to favor the development of intergranular pressure solution (Buxton and Sibley, 1981). Ebner et al. (2009a) were able to explain these observations using a simple numerical approach with a contrast in dissolution kinetics for different grains. They showed that large numbers of heterogeneities can dissolve slower than the matrix, because pinning particles accumulate along the interface, inhibiting the amplitude growth due to their slower dissolution kinetics (see Section 3).

Several authors reported that the residual can exhibit strong thickness variation along the stylolite interface (Braithwaite, 1986; Ebner 2009; Ebner et al., 2010a), including no residual at all (Heap et al., 2014). Braithwaite (1986) assumed fluid flow and thus lateral transport as a cause for this variation. Ebner et al. (2010a) discounted abrupt changes in thickness from the initial distribution of the residual material in the host rock. They found indentations in the residual where quartz-grains impinge on the stylolite (Figure 10) and proposed a mechano-chemical compaction/dissolution of the residual: heterogeneity in mechanical and dissolution properties of the grains leads to change in the amount of dissolution, and quartz-grains are associated to smaller residual thickness.

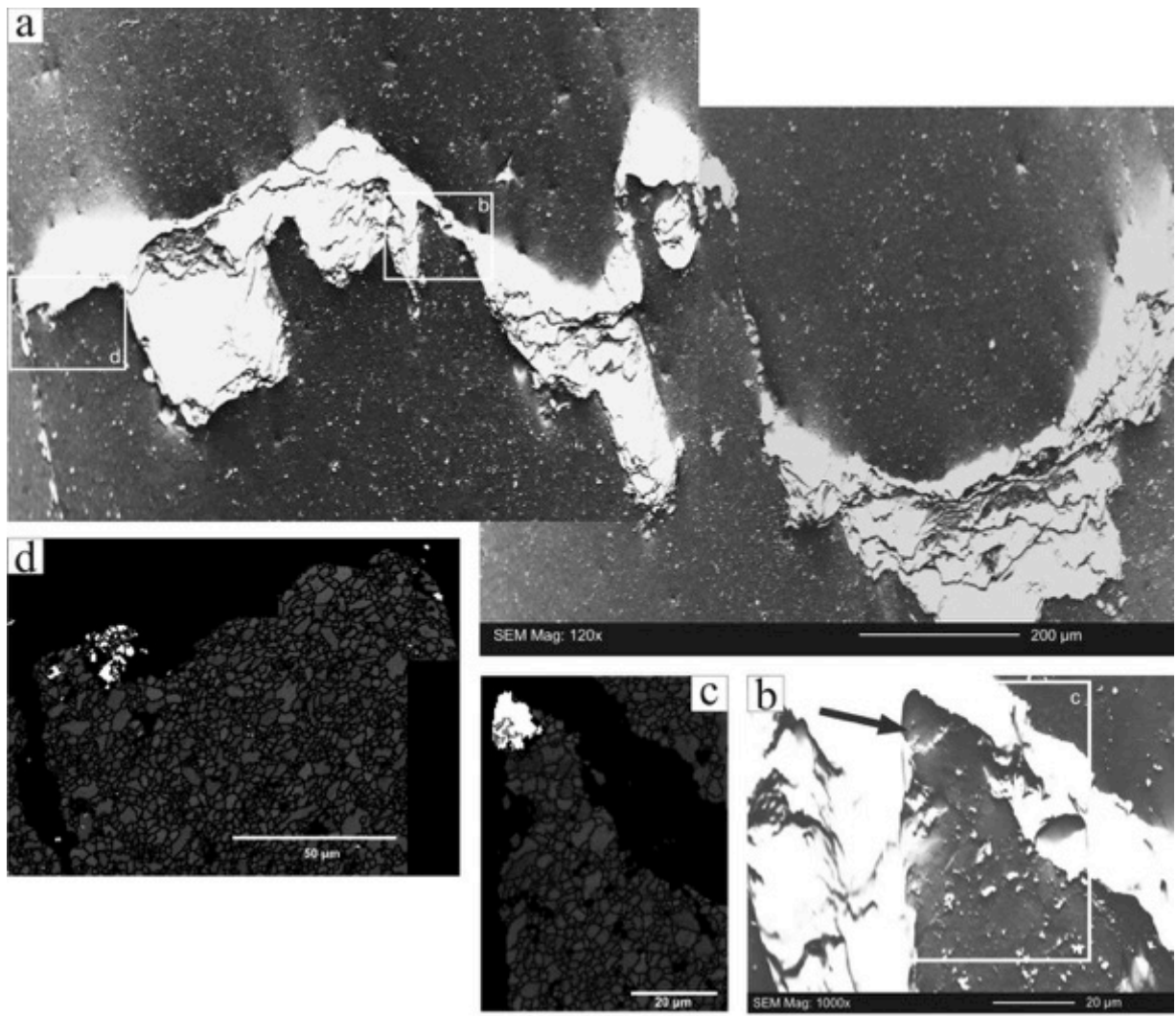

Figure 10: Scanning electron microscopy with orientation contrast images and electron back-scattered diffraction maps of mature tectonic stylolites in fine grained Jurassic limestones from SW Germany. a) Orientation-contrast image showing a mature (near horizontal) stylolite interface with significant thickness variation of the residual clay layer. Note that the topography changes from one side to the other. Frames indicated the enlargements in b and d. b) Enlargement of zone b from panel a) showing the indentation in the residual clay layer. Arrow points to a grain with a high relief. c-d) Electron back-scattered diffraction maps of the enlarged areas in panel a) and b) showing the grain outline and a color-coding for the material (dark grey – calcite; light grey – quartz; black- residual clay layer ). Quartz grains, in white on c) and d) are present along the stylolite boundary in every position displaying a large thickness variation of the residual clay layer (i.e. at the corner of the stylolite teeth, the stylolites being in black above the grains in c and d). These quartz grains are pinning grains for the morphology. Modified from Ebner et al. (2010a).

In some carbonate rocks, a gradual decrease in the amount of insoluble material transverse to the stylolite was noted while the concentration of material removed by dissolution decreased as a function of the distance from the seam center (Braithwaite, 1986). Isolated blocks of unaltered host rock may also be present in the residual due to the coalescence of individual stylolites to form stylolite networks (Ben-Itzhak et al. 2014). Detailed SEM studies by Carrio-Schaffhauser et al. (1990) and Raynaud and Carrio-Schaffhauser (1992) revealed what they called a "process zone" near stylolite tips that shows increased porosity on a micrometer scale. They argue that due to a self-localization mechanism of stylolites, pressure solution is active in a zone rather than along individual grain contacts only. Cathodoluminescene imaging revealed mineral-filled factures around grain sutures (Milliken, 1994). Indeed, a combined scanning electron microscopy-electron back scattered diffraction study of tectonic and bedding-parallel stylolites in micritic limestones showed the existence of a mm-scale altered zone around stylolites (Ebner et al., 2010a and cf Figure 10 therein). In their study, Ebner et al. (2010a), observed that in the vicinity of stylolites with a well-developed residual layer, the shape and lattice preferred orientation of matrix grains changes, and that the grain size is reduced by ~15%, leading to a porosity increase in this altered zone. Similar porosity increases in the vicinity of stylolites was also noted in several other studies (Dawson, 1988; Carozzi and von Bergen, 1987; Carrio-Schaffhauser et al., 1990; Heap et al., 2014). It can thus be concluded from these observations that although stylolites form by a localized pressure solution process, pressure solution is not confined solely to the stylolite interface and is active in a slightly wider zone (mm-scale) around the stylolite.

*2.4 Scaling laws of stylolite roughness*

The roughness of stylolites planes was analyzed using the concept of fractals, to search for scaling relationships (Karcz and Scholz, 2003; Renard et al., 2004; Schmittbuhl et al., 2004; Ben-Itzhak et al., 2012). Stylolite geometry can be measured along one-dimensional profiles in the field, two-dimensionally when surfaces are separated, or three-dimensionally via x-ray computed tomography. Their roughness, i.e. height profiles, can be characterized at all scales by means of statistical approaches used in the physics community (e.g., Santucci et al. 2010). For each point $x$ along the average profile, the height $h$ of the stylolite can be measured.

Using different transforms, for each distance $l = \Delta x$ along the profile, and perpendicular-to-plane height

$$\Delta h(x, l) = h(x + l) - h(x) \qquad (1)$$

can be defined. When $\Delta h(x, l)$ is statistically translationally invariant, such that it has no systematic trend as $x$ varies, a distribution $p_l(\Delta h)$ can be obtained for $\Delta h$ at a given difference $l$, by considering all realizations of $\Delta h(x, l)$ at all values of $x$. The profile $h(x)$ is said to be self-affine with a Hurst exponent $H$, if $p_l(\Delta h)$ is scale invariant for all zoom factors $\lambda$ under the affine transformation:

$$\begin{cases} x \to \lambda x \\ h \to \lambda^H h \end{cases} \qquad (2)$$

A common way to identify such a self-affine geometry is to obtain the characteristic out-of-plane distance, also called surface width, $w = \Delta h$ as function of the measurement window size $l$. If the profile is self-affine with a power law exponent $H$, called the Hurst exponent, this characteristic distance follows a scaling law of the type:

$$w(l) = \alpha l^H \qquad (3)$$

where $\alpha$ is a constant prefactor. If the Hurst exponent H=1.0, the interface is called self-similar and does not change characteristic shape at different scales. If the H<1, the interface is called self-affine and becomes less rough on larger scales. The classical ways (Barabasi, 1995) to extract the function $w(l)$ are:

i) A root-mean-square (RMS) average of profile width at scale $l$:

$$w(l) = \left\langle \left[ \frac{1}{l} \int_x^{x+l} dx' \left( h(x') - \frac{1}{l} \int_x^{x+l} h(x'') dx'' \right)^2 \right]^{1/2} \right\rangle \qquad (4)$$

where $\langle \, \rangle$ refers to a spatial average over $x$, the starting points of the profiles.

ii) A root-mean-square average at fixed distance lag $l$,

$$w(l) = \langle (h(x+l) - h(x))^2 \rangle^{1/2} = C_2(l). \qquad (5)$$

iii) Or more generally, one can use structure functions (Santucci et al., 2010) with higher-order moments of the distribution of height-height differences $p_l(\Delta h)$, given by:

$$w(l) = \langle |h(x+l) - h(x)|^N \rangle^{\frac{1}{N}} = C_N(l). \tag{6}$$

iv) It can as well be estimated from the "minimax" estimator, or maximum difference between points lying up to an in-plane distance $x$ from each other:

$$w(l) = \langle Max_{[x,x+l]}(h) - Min_{[x,x+l]}(h) \rangle \tag{7}$$

v) If a surface is self-affine with a Hurst exponent $H$, it also presents scaling laws for its Fourier Power Spectral Density (PSD) and Average Wavelet Coefficient (AWC) spectrum, as:

$$P(k) = \|\tilde{h}_k\|^2 \propto k^{(-1-2H)/2} \tag{8}$$

where $k$ is the wavenumber, $\tilde{h}_k$ is the Fourier transform of the profile and $P(k)$ its Fourier power spectrum, and as:

$$w(l) = \langle |w(x,l)| \rangle \propto l^{H+\frac{1}{2}} \tag{9}$$

where $w(x,l)$ is the wavelet transform of the profile (Simonsen, 1998). The fact that a profile follows a scaling law according to Eq. (4) for the root mean square, or for $C_2(l)$, or that its Fourier transform follows the scaling law Eq. (8), is often used as a weak definition of self-affinity.

For most stylolites, the height profile $h(x)$ exhibits self-affine properties over some scale range (a range of values of $l$), obeying a scaling law of the type of Eq. (3):

$$w(l) = \alpha l^H \tag{10}$$

such that the width of the surface increases as a power-law with the window of measurement $l$.

Various works have measured height fluctuations and width of numerous tectonic and sedimentary stylolites, and found that stylolites are self-affine over three different scale ranges, and are characterized by 3 main Hurst exponent regimes, depending on the scale of measurement, *l*. When roughness is measured at small scales the Hurst exponent usually is found to have a value $H = 1.1$, visible in the Fourier power spectrum (e.g. in (Gratier et al, 2005, Rolland et al., 2012)). At larger scales, usually above a scale lying between 0.05 mm to 2-3 millimeters depending on the stylolite, another self-affine scaling law is found, with a Hurst exponent $H = 0.5$ to $0.6$. This outcome was observed for numerous sedimentary stylolites, e.g. in (Renard et al., 2004; Schmittbuhl et al., 2004; Ebner et al., 2009b; Rolland et al., 2012; Rolland, 2013; Ben-Itzhak et al., 2011, 2012) (Fig. 11). The cross-over between these two scaling regimes occurs at a scale called the cross-over length *L\**.

At the largest sizes for long stylolites, above a few centimeters, another cross-over length scale appears, which we call length *X*: stylolite heights for points separated by distances *l* larger then *X* are essentially uncorrelated. This outcome leads to a different measured scaling of the surface at the largest scales, *l>X*, whereby a flattening of the scaling law towards *H=0* occurs (Ben-Itzhak et al., 2011, 2012) (Fig. 12). The length *X* at which this flattening occurs is the correlation length of the profile, beyond which deviations from a strictly flat line are uncorrelated. This last cutoff scale can be used to invert the amount of dissolution along a stylolite, even when the largest teeth have been dissolved and the height of these teeth has become shorter than the total amount of shortening (Ben-Itzhak et al., 2012).

To summarize, two characteristic length scales, *L\** and *X* can be measured on stylolites, which separates three roughness scaling regimes with typical Hurst exponents: *H* is close to 1 for the smallest scales, (Fig. 11, see also Fig. 23f) , *H* is close to 0.5 for the intermediate scales (Fig. 11 and 12), and *H* is close to zero for scales larger than *X* (Fig. 12).

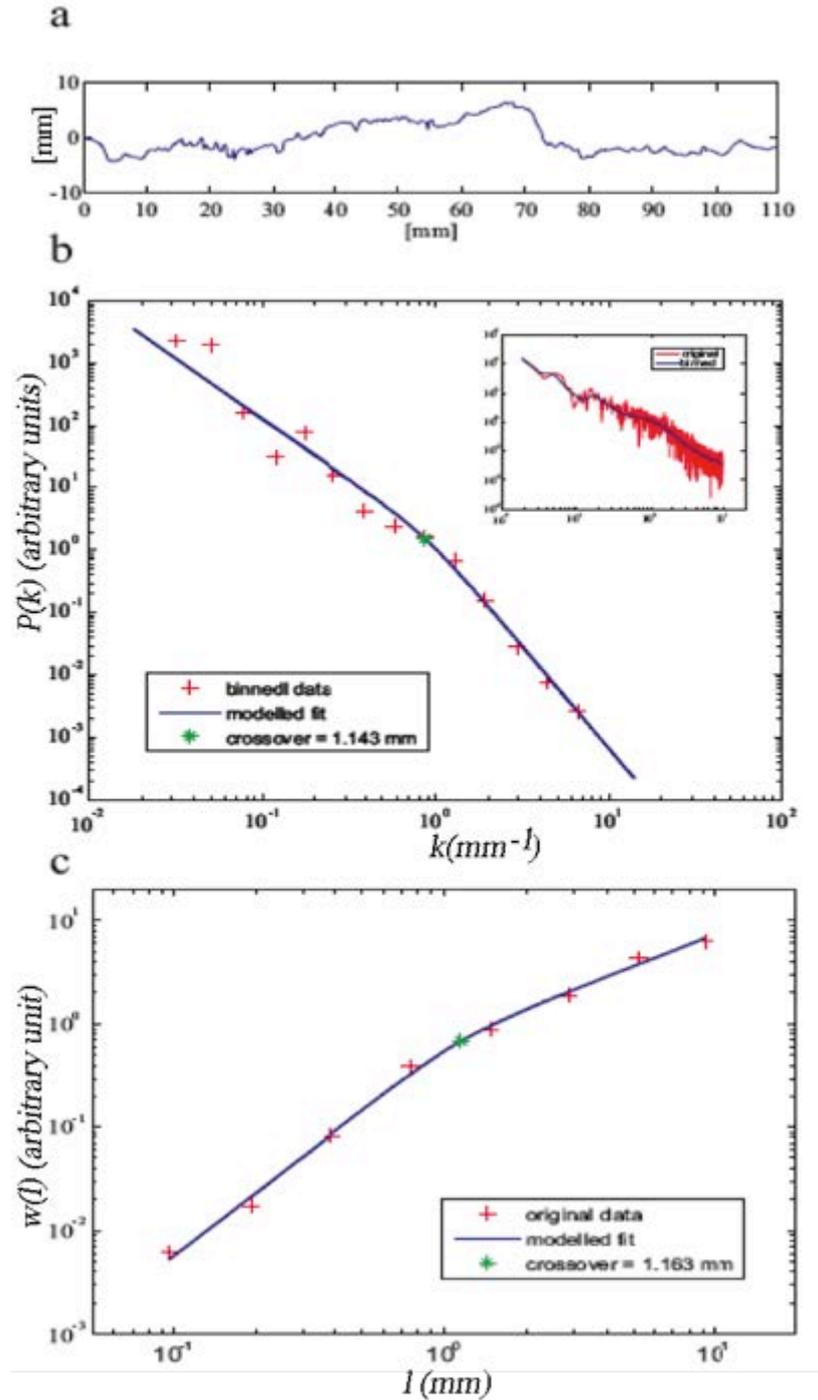

Figure 11: a) Stylolite profile extracted from a thin section on a limestone sample from Cirque de Navacelle, Southern France. b) Fourier power spectrum of the profile (see equation 8), in bi-logarithmic representation: a division into two scaling laws is seen with H=1.1 at small scale and H=0.6 at large scale. A cross-over length L* separating the two scaling laws is extracted. Length units are in mm. c) Average Wavelet Coefficient spectrum (see equation 9):

the same scaling laws and a similar cross-over length are seen using this other spectral analysis. After Ebner et al., 2009b.

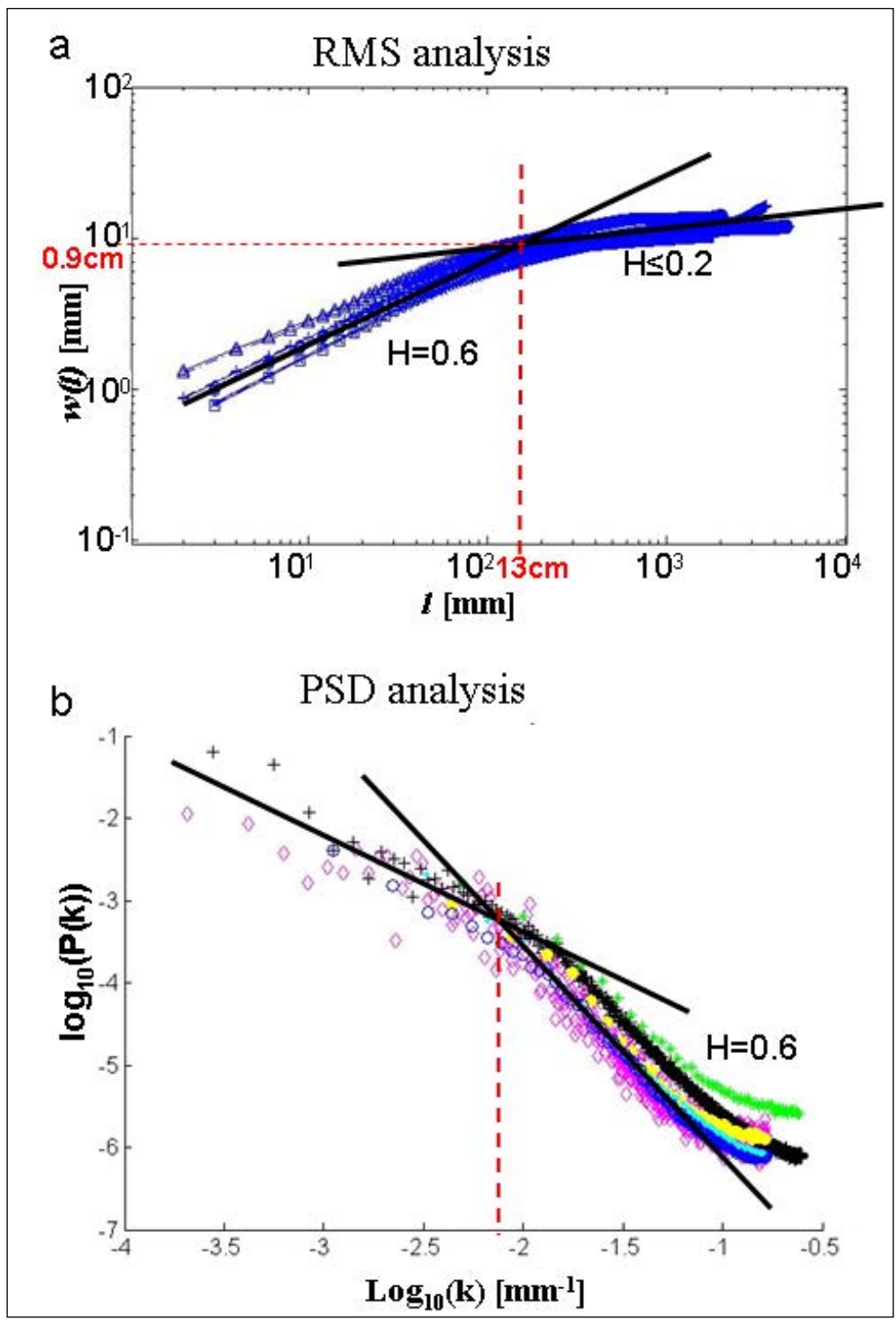

Figure 12: a) Root Mean Square and b) Fourier Power Spectral Density analysis of surface roughness of six stylolite surfaces in the Blanche cliff (Israel). In both analyses, two scaling

regimes are seen, with H~0.6 at scales up to a length scale X (here X~13cm), and a smaller slope at larger scales, l>X. is the wavenumber (modified after Ben-Itzhak et al (2012)).

In the case of tectonic stylolites, they are typically shorter than their sedimentary counterparts. Thus only the same two self-affine regimes, characteristic of small and medium sizes are found, but the cutoff length *L\** depends on the direction along the average stylolite plane,contrarily to sedimentary stylolites (Ebner et al., 2010b; Rolland et al., 2014). This direction is related to the anisotropy of the in-plane formation stress of the stylolite, for the directions along the average plane. In contrast to sedimentary stylolites, where $\sigma_2$ and $\sigma_3$ are normally found to be equal, tectonic stylolites tend to form with unequal and $\sigma_2$ and $\sigma_3$, with $\sigma_2$ defined by gravity and thus depth of formation (Figs. 13,14).

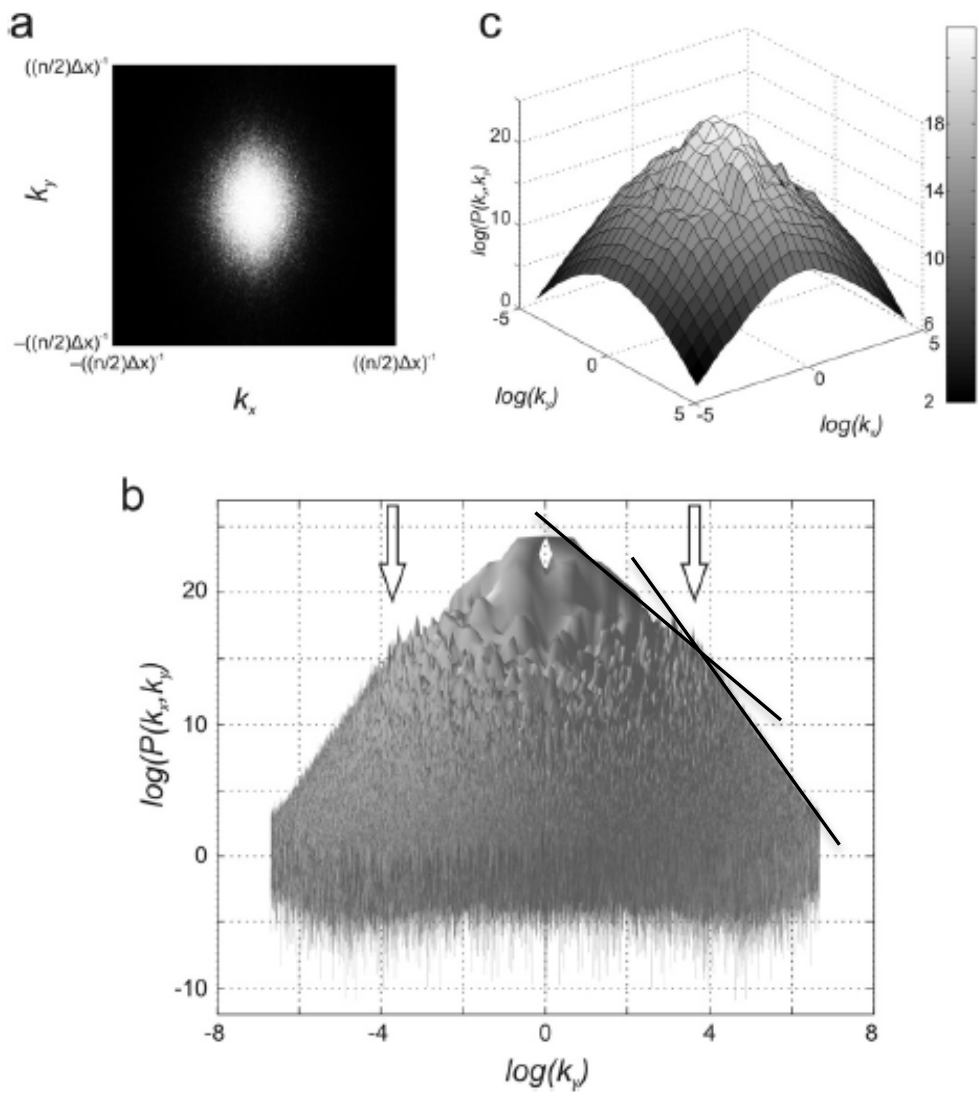

Figure 13: a) Two dimensional data analysis of a tectonic stylolite surface: the grayscale corresponds to the power spectrum of a two-dimensional Fourier transform of an open stylolite elevation as function of (kx, ky) ranging from $-n/2(\Delta x)$ to $n/2(\Delta x)$, where the data is digitized with n×n points (n=2000) with a spatial sampling distance $\Delta x = 25$ μm. The directions (x, y) coincide with the horizontal and vertical directions along the stylolite, respectively. The anisotropy observed is also seen in the bi-logarithmic representation on b) and c). The oblique view of the 3D surface representation shows that the two small-scale and one medium-scale power laws, corresponding to straight lines in this representation, are also present for such tectonic stylolite. After Ebner (2010b).

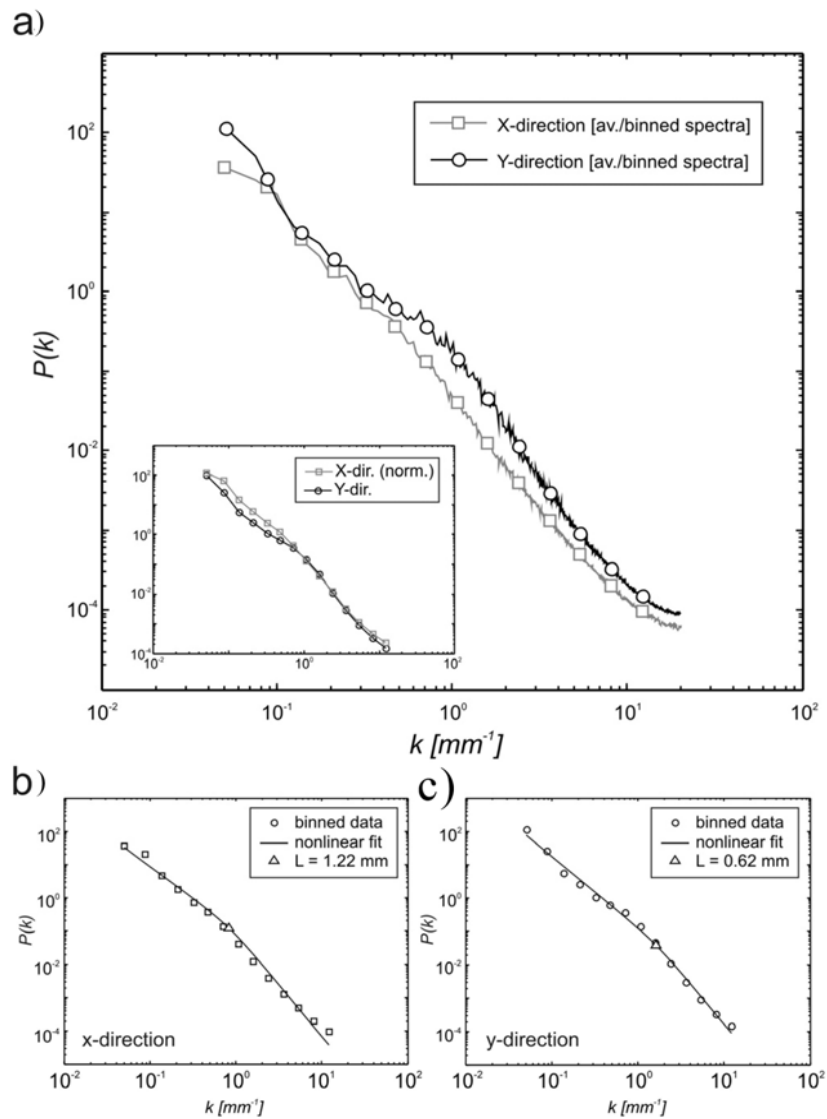

Figure 14: One-dimensional analysis of profiles displayed with a two-dimensional representation in Fig. 13. a) Power Spectral Densities for the two perpendicular directions of the tectonic stylolite coincide at small scales (large k). They differ at small k (large scale), and also they follow self-affine scaling laws with the same exponent – H=0.5 at large scale, and H=1.1 at small scale. The prefactors of the large scale scaling laws $P(k) \propto k^{-1-2H}$ differ along the two directions. b) Power Spectral Density along the x-direction, and non-linear fit identifying the small-scale power-law (straight lines), the large-scale one, and the cross-over length (triangle). c) Power Spectral Density along the y-direction, nonlinear fit and cross-over length. The cross-over lengths differ along both directions, with L*=1.22 mm along the x-direction, and L*=0.62 mm along the y-direction. After Ebner (2010b).

To conclude, these fractal analyses of stylolite roughness show the existence of cross-over length scales and well-identified fractal exponents. These data indicate that several physical processes have been responsible for such geometrical properties. They are used to constrain physical models of stylolite formation, as described in Section 4.

*2.5 The continuity from stylolites to spaced solution cleavage*

Similar to stylolites, solution seams of spaced pressure-solution cleavage are surfaces of localized dissolution. The main difference is that solution seams are smooth in contrast to the rough stylolites. Both structures can develop simultaneously, for example when stylolites and solution cleavage are associated with the folding of limestone and marl layers (both are parallel to each other in Figure 15, a-b). Spatially, stylolites can evolve into solution cleavage in lateral continuity (fold hinge in Figure 15 c-f), with cleavage refraction being common between competent (limestone) and incompetent (marl) layers (Ramsay, 1967). The spatial density of dissolution seams varies in relation to the lithology of the deformed rock (Fig. 15): (i) only one dissolution (stylolitic) surface in pure limestone (100% of calcite) (Fig. 15 a-b); (ii) a very dense array of dissolution seams (solution cleavage) in marl (mixture of about half calcite and half phyllosilicates content) (Fig. 15 a, b) and an evolution from solution cleavage in marl to stylolites in limestone (Fig. 15 c-f). Due to the accumulation of strain through material dissolution, stylolite density increases with strain.

Contrary to stylolites that develop mainly in structurally heterogeneous granular rocks (i.e. large-grained ones), solution cleavage develops in more structurally homogeneous polymineralic fine-grained rocks (Hobbs et al., 1976; Ramsay, 1967). This induces several differences: (i) fractures oriented perpendicularly to stylolites that often bound the stylolite peaks in limestone do not develop so easily in marls; (ii) heterogeneities (initial or induced by the deformation such as local porosity changes) are larger in limestone or in other monomineralic aggregates than in marls; (iii) the density of heterogeneous sites on which dissolution seams may initiate is much greater in marls than in limestone; (iv) when the insoluble mineral content is high enough, their passive concentration in the solution seams acts as a "smoothing" process as deformation increases. Consequently, there is no gap between the two types of structure but rather a continuous change (Figure 15).

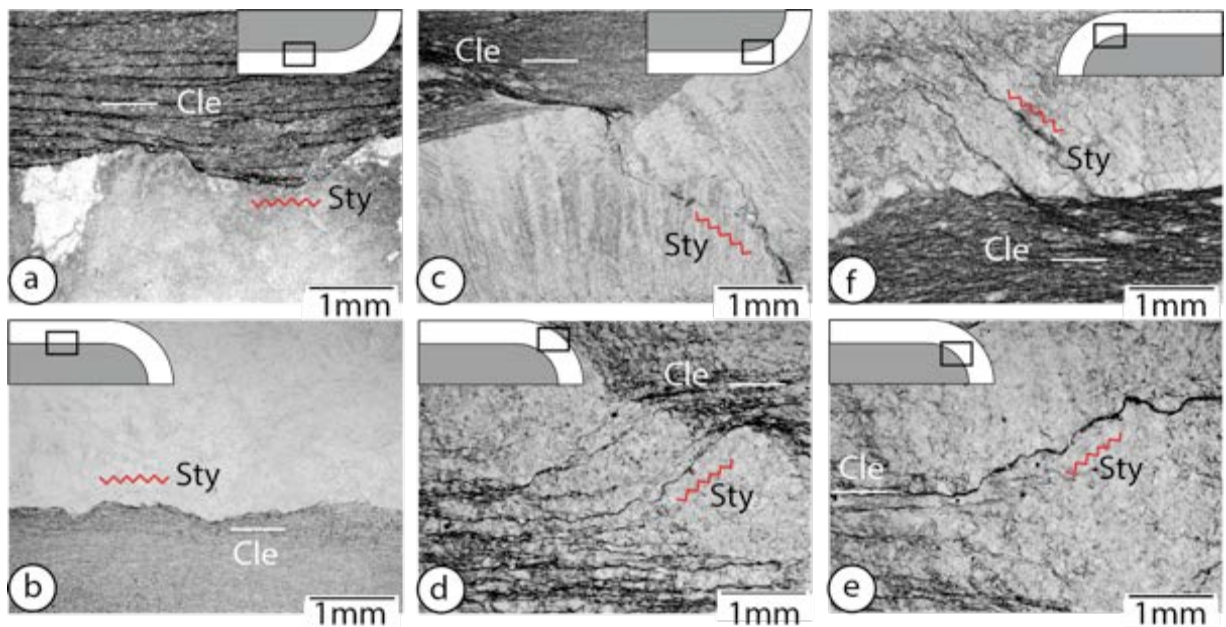

Figure 15: Spatial transition from stylolites to solution cleavage in competent and incompetent layer respectively: the four samples are marls and limestones from the folded sedimentary basins of the Western Alps. Insert shows sample location in host fold. Cle=solution cleavage, Sty=Stylolites.

## 3. Thermodynamics and kinetics of stylolites: why do they form?

*3.1 Driving forces*

Stylolite formation is commonly attributed to a localization of the dissolution-re-precipitation process called pressure solution (Tada and Siever, 1989 and references therein). Pressure-solution creep is broadly defined as dissolution and re-precipitation driven by spatial variations in chemical potential along grain surfaces or across larger regions where solution occurs at locations of high chemical potential, the dissolved material is transported through the fluid phase, and precipitates in regions where the chemical potential is less (e.g., Gratier et al., 2013a). This approach is based on the theory of non-equilibrium thermodynamics (Gibbs, 1877). It is probably not surprising that many parameters are observed to affect stylolite formation, because the chemical potential is controlled by stress, plastic and elastic strain energies, crystal orientation and interface curvature (Deboer, 1977; Lehner, 1995; Paterson, 1995; Shimizu, 1995). Chemical potential includes an electrochemical potential (Greene et al., 2009) where spatial variations in electric charge at the surface of minerals may arise as well, often due the presence of minerals such as clays (Renard et al., 1997).

The chemical potential difference between different locations, $\Delta\mu$, is given by (Shimizu 1995; Renard et al. 1999; Røyne 2012)

$$\Delta\mu = \Omega\Delta\sigma_n + \Delta f_s; \quad \Delta f_s = \Delta f_{el} + \Delta f_d + \Delta f_{se} + \Delta f_{sc} \quad (11)$$

Here $f_s$ is the molar Helmholtz free energy, $\sigma_n$ is the normal stress acting on the grain surface and $\Omega$ is the molar volume of the mineral. The Helmholtz free energy arises from various contributions: $f_{el}$ is the contribution from the *elastic* strain energy; $f_d$ is the *plastic* strain energy stored in defects (e.g. dislocations); $f_{se}$ is the contribution from surface energy; and $f_{sc}$ is a contribution due to effects from surface charges and ionic environment on the local solubility and strength of intermolecular mineral bonds. Spatial variations in normal stress ($\Delta\sigma_n$) and Helmholtz free energy ($\Delta f_s$) arise along the grain surface, between the stressed contact and the free surface of grain, and on smaller lengthscale within the contact itself, due to roughness and heterogeneity. The relative values of the different forcing terms in equation 11 provided for quartz and calcite in Table 1.

The first term is often considered the largest and includes the difference between normal stress on grain contacts, $\sigma_{gb}$, and the fluid pressure acting on free faces of grains, P, often assumed to be hydrostatic. The molar volumes $\Omega$ for calcite and quartz are $3.7 \times 10^{-5}$ m³/mol and $2.3 \times 10^{-5}$ m³/mol, respectively. The other terms, arising from Helmholtz free energy differences, are usually considered small, but their relative magnitude strongly depends on the specific system.

The elastic strain energy is calculated using $\Delta f_{el} = \frac{1}{2}\Omega(\sigma_{gb}-P)^2/E$, where E is the Young's modulus of the mineral, ranging between 70-100 GPa for calcite and quartz. This term is ~$10^{-3}$ times smaller than the 1$^{st}$ term (assuming $\sigma_{gb}$=100 MPa), and hence, makes a negligible contribution to $\Delta\mu$.

The plastic strain energy contibution, $\Delta f_d$, is more complex to calculate and is highly variable: e.g. damaged rock in a fault zone may have large plastic strain, while far away from a fault zone this term may be small. Plastic strain energy is estimated using two main methods: (i) estimating density of dislocations (e.g. Schott et al., 1989) and (ii) estimating partition of plastic work between heat generation and cold-work (i.e. defects) (e.g. Shimizu, 1995), using:

$$\Delta f_d = \Omega \alpha \varepsilon_p \sigma_{gb} \qquad (12)$$

where $\varepsilon_p$ is the plastic strain and $\alpha$ is the non-dissipated fraction of the plastic external work performed on the system (i.e. fraction of cold-work). The values of $\varepsilon_p$ and $\alpha$ vary widely depending on material, strain, strain rate, and tectonic history of rocks, and have values up to $\alpha$=0.5 and $\varepsilon_p$=0.5 (Shimizu, 1995; Austin and Evans, 2009; Rosakis et al., 2000; Benzerga et al., 2005; Paterson, 1995; Cheng and Cheng, 2004; Ben-Itzhak, 2016). Using these upper bounds in Eq (12) predicts that the term $\Delta f_d$ can reach 25% of the 1st (stress) term in Eq (11) (see Table 1). In contrast to these predictions of $\Delta f_d$ using the work-partition method, the dislocation method for estimating $\Delta f_d$ suggests that plastic strain, even in highly stressed locations, provides a negligible contribution to free-energy (e.g Bisschop, et al. 2006.) Bridging predictions from these two methods is not straightforward, although progress has been done regarding this (Benzerga et al., 2005).

To estimate the surface energy term, we use $\Delta f_{se}=2\gamma\Delta k\Omega$, where $\Delta k$ is the variation in surface curvature and $\gamma$ the surface free energy. The surface curvature, $\Delta k$, increases with decreasing length scale of roughness, r, scaling as $1/r$. Although grain sizes in undeformed rocks have length-scale r > 1μm, roughness scale in deformed and damaged rocks is sub-micronic, as observed both in indentation experiments (e.g. Fig. 7 in Ben-Itzhak, 2016; Figs. 5 and 6 in Croize et al., 2010; Omori et al., 2015) and in natural faults (e.g. Chester et al., 2005; Siman-Tov et al., 2013). These observations suggest that small-scale roughness, r, may range between 10 nm-1000 μm, depending on grain size, micro-crack existence, and roughness of grains, from which we calculate that $\Delta k$ ranges between $10^3$-$10^8$ m$^{-1}$, depending on conditions. The surface energy, $\gamma$, ranges for calcite between 0.15-0.6 J/m$^2$ (Royne et al. (2011)) depending on the chemical activity of water, and for quartz γ~0.35 J/m$^2$. Table 1 shows a range of values for surface energy, with large values corresponding to intensely damaged rocks with nano-scale roughness, and small values assuming curvature dictated by grain size.

The last term, $\Delta f_{sc}$, the surface-charge contribution, may arise from clays or dissimilar materials lining grain surfaces, but is difficult to estimate. Moreover, for this last term, a debate continues as to whether it must be included directly in the chemical potential, or if it is responsible only for a kinetics effect (see Section 3 in Gratier et al., 2013a). We do not add its contribution to Table 1 due to a lack of a general understanding about its role.

| Chemical potential difference term (J/mol) | Quartz, 300 MPa, 400°C, grain size 0.1 mm | Calcite, 10 MPa, 25°C, grain size 5 micron | Calcite, 100 MPa, 25°C, grain size 5 micron |
|---|---|---|---|
| Normal stress | 6900 | 370 | 3700 |
| Plastic strain energy | < 1700 | < 90 | <900 |
| Surface energy | 0.2-1600 | 2-4400 | 2-4400 |

Table 1: Orders of magnitude of the energy contributions for Eq. 11 with quartz and calcite. For all calculations, stress values mean $\sigma_{gb}$ - P.

Although pressure solution derives its name from the assumed mechanism controlling this process, i.e. normal stress (1$^{st}$ term in Eq. 11), Table 1 suggests that other terms in the chemical potential may drive or contribute to mass transfer creep, depending on conditions such as ambient stress, plastic damage and deformation, mineral assemblage, and importantly grain size and roughness. For example, mass transfer creep in shallowly buried fine-grained rocks (i.e. under low $\sigma_{gb}$) may be controlled by surface energy terms and not by stress, as the experiments of Visser et al. (2012) demonstrated. Meike and Wenk (1988) observed that plastic deformation (dislocation concentrations) may drive pressure solution along solution seams. Ben-Itzhak et al. (2016) showed that roughness and possibly plastic strain may control the precipitation step for pressure-solution operating in damaged calcite. Evidence that other driving forces are important, in addition to stress has also been provided from field work, where several workers (Dunnington, 1967; Bathurst, 1971; Buxton and Sibley, 1981; Tada and Siever, 1989, Gruzman, 1997) have suggested that stylolites in limestones can form very early in the diagenetic process, occurring at burial depths less than 100 m, where the normal stress is very small. In addition, in some situations, a steady-state pressure solution rate is reached when different pressure solution mechanisms occur simultanuously and interact with each other. Such a situation was observed by Karcz et al. (2006, 2008) in pressure-solution of halite, where plastic deformation occurred together with thin-film pressure-solution and strain-energy-driven pressure solution, to produce a steady pressure-solution convergence rate in a channel-and-island structure. Thus, pressure-solution is likely dominated by different processes, and possibly several processes simultanously, at different development stages depending on the environment.

Since the dominant driving control may depend on the particular system, one may ask whether the driving force for dissolution on stylolites differs in any way from that of intergranular pressure solution. A main factor for stylolites may be the effect of the relatively thick and continuous lining of minerals such as clays, phyllosilicates, pyrite, and organic matter, along these surfaces. The contact of calcite or quartz with this lining may change or enhance either driving forces or kinetics (Hickman and Evans, 1995; Bjorkum, 1996; Renard et al., 2001; Renard et al., 1997; Hickman and Evans, 1991; Alcantar et al., 2003; Meyer et al., 2006; Greene et al., 2009; Kristiansen et al., 2011). How clays, phyllosilicates, oxides and organic matter enhance pressure solution is not fully understood, and suggestions include enhancement of kinetics by retaining open fluid pathways (Heald, 1956,1959; Weyl, 1959; Wanless, 1979; Marshak and Engelder, 1985; Dewers and Ortoleva, 1990; Hickman and Evans, 1995), changing reactions by varying pH (Thomson, 1959), electrochemical gradients amplified by clay electric charge effects (Walderhaug et al., 2006; Anzalone et al., 2006), creating soft inclusions that concentrate stress (Gratier et al., 2015), or a combination of these factors. Modeling results suggest that the stress driving force alone is insufficient to drive stylolite elongation (Aharonov and Katsman 2009), although it was found necessary to induce roughening (Koehn et al., 2007, 2012; Rolland et al., 2012).

*3.2. Limiting processes*

In addition to estimating the driving forces for the dissolution and reprecipitation creep, it is important to consider what limits the reaction rate. These factors could be the dissolution rate, the precipitation rate or the diffusion rate from the dissolution site to the precipitation site. The slowest step of this serial process controls the rate of the entire process, which is said to be diffusion-limited if this step is the diffusion, or reaction-limited if the slowest step is the precipitation or dissolution reaction. Several pressure-solution creep laws are possible depending on the limiting step (e.g., see discussion in Gratier et al., 2013a).

The limiting process can change with stress, grain size, and other factors (Bernabé and Evans, 2014). For example, the model of Renard et al. (1999) predicted for sandstone, that diffusion-limited dissolution at the grain contacts dominates pressure solution for small grains, yet for larger grains or at greater depth, diffusion within thin fluid films is considerably slowed, to the point when other mechanisms start to dominate the overall pressure solution rates, such as precipitation rates, plastic strain energy and surface energy.

Since calcite has orders of magnitude faster reaction rates than quartz, the slowest step for pressure-solution of calcite is often the diffusion, causing pressure-solution in calcite in many situations to be is diffusion-limited rather than reaction limited, more often than for quartz. The limiting step in calcite is still less understood and more variable than in quartz. It depends strongly on the exact conditions, notably temperature: While some experiments suggest that deformation in calcite is controlled by diffusion (Zhang and Spiers, 2005a; Zubtsov et al., 2005), other results suggest that the rate may be controlled by precipitation kinetics (Baker et al., 1980; Zhang et al., 2002; Zhang and Spiers, 2005b). The lack of consensus on the rate-limiting process of pressure solution in carbonates is partially due to the absence of good agreement between macroscopic strain-rate laws and experimental results, which show greatly variable of strain rates as function of stress (e.g., the four orders of magnitude change in strain rate for the same applied stress, shown in Fig. 12 of Croizé et al. (2010)). Croizé et al. (2010) attributed variations in pressure solution rates observed in their nano-pressure solution experiments to presence or absence of crack growth together with pressure-solution, because they observed that when cracks grew, strain rates increased by orders of magnitude. The observed cracks in this case penetrated through the dissolving surface, radially emanating from the indenter, and having tracelengths of a few tens of microns. In general, microcracks are often observed around pressure solution features (e.g. Milliken, 1994; Dickinson and Milliken, 1995; Land and Milliken, 2000; Makowitz and Milliken 2003), which suggests that brittle processes could play an active role in enhancing pressure solution. Two main suggestions are offered to explain how cracks enhance pressure solution: i) fractures short-circuit fluid pathways, enhancing pressure solution if diffusion is rate limiting (Gratz, 1991; Gratier et al., 2014), and ii) new fractures add highly reactive surfaces, and plastic defects, that both enhance surface energy, plastic strain energy, and provide new locations for dissolution /precipitation (Ben-Itzhak et al. 2016).

Turning from pressure solution in general to stylolites in particular, it is clear that material transport away from the dissolving stylolite surface, and the location of the sink, are central to the process, and often limit it. Field evidence suggests that precipitation from dissolving stylolites occurs in close proximity to stylolites (Raynaud & Carrio-Schaffhauser, 1992; Ben-Itzhak et al., 2012; Emmanuel et al., 2010), and suggests also that dissolution on stylolites may shut down when no place exists for precipitation (Ben-Itzhak et al., 2012). Angheluta et al. (2012) modeled dissolution on stylolites coupled with diffusion into surrounding pores and precipitation that clogs the pores along the diffusive path, to predict dissolution rates on stylolites. The results of this model (Angheluta et al., 2012) predict very low rates of dissolution along stylolites, that are about a million times slower than the rate constants for mineral-interface dissolution measured in laboratory experiments, since the dissolution process along the surface is calculated to be controlled by diffusion and precipitation, and not by dissolution rates. However, this model considers a situation where all dissolved material from a stylolite is transported through pores, and precipitates into a fixed pore volume without generation of veins. In reality, pressure solution on stylolites (and also on single grains) is often observed to be coupled with the evolution of fractures and veins (e.g. Alvarez et al., 1976; Rye and Bradbury, 1988; Willemse et al., 1997; Renard et al., 2000; Smith 2000; Croizé et al., 2010; Gratier, 2011; Gratier et al., 2014; Katsman, 2010; Ben-Itzhak et al., 2014). Stylolites and veins often create a fully connected network (Smith, 2000; Ben-Itzhak et al., 2014) where the veins have a role in allowing fast diffusion, providing precipitation locations, and hence, enhancing dissolution by stylolites (e.g., Fig. 7b, 16). Fig. 16 illustrates the typical crossing between horizontal sedimentary stylolites and quasi-vertical veins cutting through. Fig. 7b illustrates how the termination of stylolites is often the beginning of quasi normal-to-stylolites veins. Figs. 2, 9, 10 and 16 also show that stylolites and veins are the location of precipation of secondary phases.

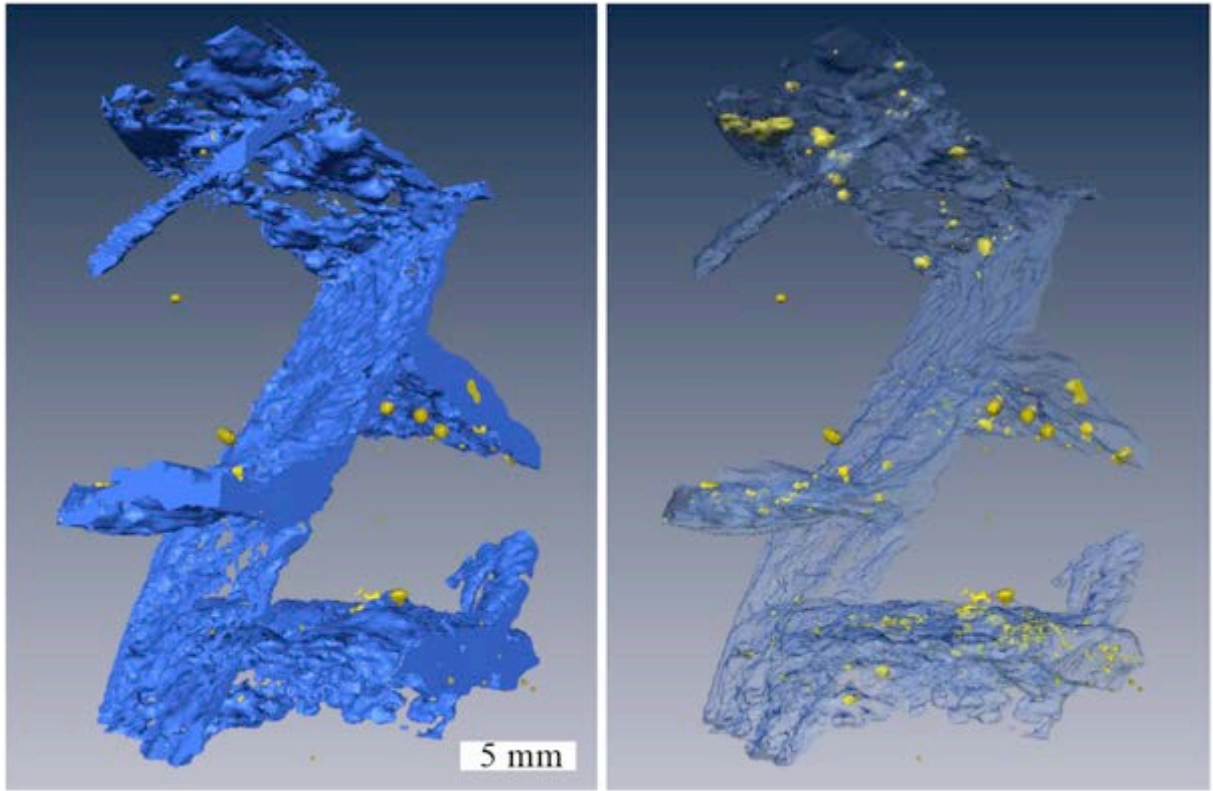

Figure 16 : Computerized tomographic image of a set of stylolites and veins: the blue color is inside the horizontal stylolites and quasi vertical fractures, the yellow color corresponds to ore minerals. The left panel displays the stylolites and fracture in opaque blue, underlining their structure, and in translucid blue on the right panel, allowing to better visualize the embedded ore minerals.

*3.3. Controlling factors in stylolite formation*

Many observations suggest that stress plays a major role in stylolite formation: stylolitic teeth direction is parallel to maximum compression (e.g. Stockdale, 1922; Geiser and Sansone, 1981; Rispoli 1981; Eyal and Reches, 1983); tectonic stylolites tend to occur in regions of high compression, perpendicular to maximum compression (Fletcher and Pollard, 1981; Rispoli, 1981); and the transition in fractal scaling of roughness correlates with stress (Schmittbuhl et al., 2004; Ebner et al., 2009b). In compressive regions, increasing numbers of cleavage surfaces form with increasing strain, new surfaces forming in between older ones, with apparently no underlying structures pinning these locations (Alvarez et al. 1978). These new surfaces can potentially become stylolite nucleation locations.

In contrast, other observations suggest that high stress is not the only controlling factor and is possibly not even necessary for stylolite formation. A large body of evidence suggests an equally important role for compositional heterogeinity for determining stylolite location and extent. Sedimentary stylolites often form along bedding planes, some forming even at very shallow burial depths of tens of meters while experiencing extremely low stress (Tada and Siever, 1989; Engelhardt 1960). They nucleate on pre-existing structures (Stockdale, 1922), especially bed surfaces (Tondi et al., 2006; Ben-Itzhak et al., 2012; Rustichelli et al., 2015) or joints (Geiser and Sansone, 1981). Both textural heterogeneities, such as sharp changes in grain size and texture, and mineralogical heterogeneities, such as clays, have been found to control stylolite localization (Rustichelli et al., 2012; Rustichelli et al., 2015). A similar impact of structural heterogeneity was also observed for the development of compaction bands in porous carbonates (Cilona et al., 2012, 2014). Heald (1955, 1956, 1959) suggested that the original distribution of clay determines the location of pressure solution, so that stylolites will develop where clay is concentrated in clay partings. Consequently, uniformly distributed clays may on contrary lead to pervasive intergranular pressure solution or spaced solution cleavage instead of stylolites (Fig. 15). This suggestion is supported by observations of stylolites changing laterally into clay layers or partings (Bushinskiy, 1961; Park and Schot, 1968; Wanless, 1979). Similar conclusions were reached by Bjorkum (1996) in sandstones about the role of mica promoting pressure solution. The role of heterogeneity in determining stylolite locations is further attested by deep-sea drill cores that show stylolites progressively developing from flasers as depth increases (Lind, 1993), although stylolites are relatively rare in deep basins, and lack increase in abundance with depth; and by the juxtapositioning of different geometries of stylolite (Ben-Itzhak et al., 2014). Much field evidence points toward the enhancing effect of clay on pressure solution localization and stylolite development (Engelder and Marshak, 1985; Heald, 1955, 1956; Marshak and Engelder, 1985), although it was suggested that perhaps only certain types of clays (Walderhaug et al., 2006) or certain concentrations induce stylolitization.

In terms of the controlling factors for dissolution rates on stylolites, observations suggest that stylolite formation is transport-controlled, and that in the extreme case, stylolites will stop dissolving once the area around them becomes clogged by precipitates (Stockdale, 1922; Ben-Itzhak et al. 2012). Data of porosity as function of distance away from stylolites (Emmanuel et al., 2010) is well fitted by a model (Angheluta et al., 2012) that simulates a combined process of dissolution-diffusion-precipitation (Fig. 17).

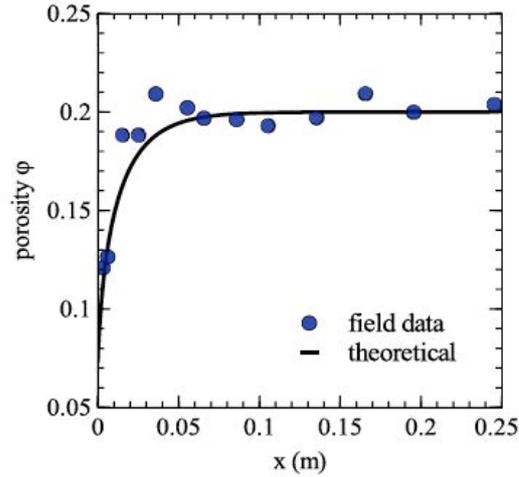

Figure 17: Comparison between field measurement of the porosity profile in sandstone, as function of distance, x, from stylolite surface, (Emmanuel et al., 2010) and the theoretical prediction for this porosity, after the model of Angheluta et al., (2012). The porosity is small near the stylolite due to precipitation clogging pores, and approaches the background value away from it.

This model predicts that for realistic conditions, dissolution on stylolites occurs in a diffusion-precipitation limited regime. The rates of dissolution on a stylolite predicted by the diffusion-precipitation controlled model of Angheluta et al. (2012) agree well with the measured sizes of stylolite's teeth. It predicts that dissolution on stylolites occurs quite sluggishly, such that quartz dissolves at a rate of $V_{sty}= 10^{-4}$ to $10^{-3}$ m/Myr, and $V_{sty}=$ 0.001 to 0.1 m/Myr for calcite.

Taking the Stø formation as an example, which corresponds to siliciclastic rocks that formed during the lower Jurassic, about170–190 Myr ago, cores of stylolites extracted from this formation show teeth of a few centimeters in amplitude (Walderhaug et al., 2006). In this formation, Walderhaug and Bjorkum (2003) studied how quartz cementation and porosity varied as function of the distance to stylolites, and attributed large porosities away from stylolites to the low concentration of stylolitic precursors. Assuming dissolution magnitude on each stylolite to be L=0.05 m, Angheluta et al. (2012) predicted that the time for its development is $t_d = L/V_{sty}$. Using $V_{sty}$ for quartz as $10^{-3}$ m/Myr, they concluded that the Stø stylolites developed over $t_d$ =50 million years. The prediction of such long time scales for quartz may also explain why stylolites in quartz are mainly found at depths exceeding 1.5 km burial (Tada and Siever, 1989). Observable teeth (i.e. L ~ 0.01 m) develop over at least 10 Myr, and, assuming a burial rate of 10 Myr/km, it provides an estimate for a minimum observed depth for initial stylolitization in siliciclastic rocks. Using directly the thickness of this formation (around 100 m (more precisely from 30 to 150 m (Halland et al. 2011)), and its age (170 to 180 Myr ago), leads to an estimate of burial rate around 100 Myr/km, and thus a lower estimate of depth at which such developed stylolites can be observed - 50Myr/(100 Myr/km) = 0.5 km. The burial rate is actually between these two values, 10 Myr/km and 100 Myr/km, because this last estimate does not account for later compaction or erosion. Given this model, once stylolites are mature enough to be observed (at some minimum burial depth, that depends on the dissolution kinetics), they grow, without necessarily becoming more abundant, until at some greater burial depth, water ceases to flow efficiently and the chemical reactions slow down or stop.

In contrast, stylolites in calcite are observed to form as shallow as tens of meters (Tada and Siever, 1989). Using $V_{sty}$ of 0.01-0.1 m/Myr for carbonates, the time to grow a stylolite with L = 1 cm teeth is predicted to take typically $t_d \approx$ 0.1-1 Myr in carbonates. Assuming a burial rate of about 10 Myr/km, it follows that sediments can be buried down to 10 meters in 100 Kyr, which implies that stylolites in carbonates can be observed at shallow burial depth (10 to 100 m). This calculation suggests that the minimum depth below which stylolites may develop may be controlled by the duration of formation, $t_d$, and not by the overburden stress as usually assumed.

A continuous yet slow formation, (set by the slow transport, and not by reaction rates), also agrees with observations in Ocean Drilling Project cores where pelagic sediments (chalks) show that stylolites develop continuously with depth (Lind, 1993) – although they are not so common in these cores. Many stylolites were observed in cores, at depths from 150 to 800 m through the Dogger and Oxfordian calcite formations of Bure-sur-Meuse (Rolland et al., 2014).

## 4. Understanding stylolites, their formation and scaling through numerical modeling

In this section, we introduce models for studying the different aspects of stylolites formation and evolution. We also discuss the scaling of stylolites and stylolite networks, and how this characteristic can be extracted from field data. We show that the physics-based models enable the interpretation of quantitative characteristics of the scaling laws for the morphology of stylolites, and that they can be used to constrain formation conditions (stress, strain, displacement, initial conditions, chemical heterogeneity).

In comparison to the large body of field studies investigating parameters controlling stylolite formation, physical and mathematical models of stylolite formation are rather scarce. The majority of the existing models deal with mechanisms controlling the roughening of the stylolitic surface (Guzzetta, 1984; Railsback, 1998; Gal et al., 1998; Schmittbuhl et al., 2004; Koehn et al., 2007; Bonnetier et al., 2009; Ebner et al., 2009a,b; Rolland et al., 2012; Gal and Nur, 1998; Gal et al., 1998), whereas only a few deal with how the surface of the stylolite, develops in an initially stylolite-free rock (Fletcher and Pollard, 1981; Merino et al., 1983; Dewers and Ortoleva, 1990; Aharonov and Katsman, 2009). Even fewer works deal with the formation of stylolites networks (Fautin and Robin, 1992, 2002; Ben-Itzhak et al., 2014). The scarcity of work dealing with genesis of stylolites is perhaps due to the fact that stylolites have never been produced in the laboratory, except on the sub-grain scale (Gratier et al., 2005) and as embryonic solution seams (Gratier et al., 2015), so the physics of their evolution is thus not well constrained experimentally. In addition, the lack of strong patterns for where stylolites can be found in deep sedimentary basins make the system difficult to systematize. Their formation depends on some combination of stress and temperature history, initial rock composition, presence of fluids, and the distribution of heterogeneities.

Stylolites are often described as isolated and spatially limited surfaces (e.g., Fletcher and Pollard 1981; Stockdale, 1922). Yet, in the field they are more likely found in close proximity to other stylolites or other structures, primarily Mode I and Mode II fractures (Peacock and Sanderson, 1995; Smith, 2000; Ben-Itzhak et al. 2014). The fact that they appear in groups may be associated to particular initial rock composition and details of the burial history, that can for example lead to fracturing, enhancing fluid transport. In addition, this could be related to a positive feedback loop in stylolite formation. Stylolite populations may be categorized as either isolated, long sub-parallel stylolites, or anastamosing networks, with or without connections via fractures (Ben-Itzhak et al., 2014; Kaduri, 2013). Based on these observed morphologies, formation models divide into those that simulate isolated stylolite initiation, roughening and lengthening, and those that simulate the initiation of stylolitic populations, specifically formation of long sub-parallel stylolites, including the spacing between them, and the evolution into anastamosing networks.

*4.1 Types of models for stylolites*

Modelling of stylolitic initiation has led to a discussion about the importance of different driving forces for the localization of dissolution planes in rocks. To model the styolite development, one has to treat the deformation of the solid and link this deformation with the dissolution of matter, coupled with transport and possibly precipitation. Since stylolites are not found everywhere in buried rocks, initiating their formation requires conditions and heterogeneities (e.g., stress conditions, fluid characteristics, host material composition, grain size distribution, porosity structure, etc.) that favor the localization, followed by roughening of a stylolite interface.

The dissolution of material at a stylolitic interface is a function of differences in Helmholtz free energies, as well as differences in normal stress at the interface and can be described as:

$$D_i = k_i V_s \left(1 - exp\left(\frac{-\Delta\sigma_n V_s - \Delta\psi_s}{RT}\right)\right) \qquad (13)$$

where $D_i$ is the dissolution rate of the interface, $k_i$ is a rate constant that varies depending on the mineral, $V_s$ is the molecular volume of the solid, $\Delta\sigma_n$ represents the changes in normal stress along the interface with respect to the average one, and $\Delta\psi_s$ the change in Helmholtz free energy (a function of elastic, plastic and surface energy), *R is* the universal gas constant, and *T* the temperature. If we assume that the fluid is saturated and at chemical equilibrium with the solid submitted to the average overall stress in the system, then the driving force for the normal stress term is only the stress difference along the interface and is as small as the Helmholtz free energy (Gal et al., 1998). However, if one considers that the normal stress difference is the difference between a stressed and an unstressed surface, the normal stress term becomes dominant. For example, Koehn et al. (2007, 2012) found that models for stylolite roughness only lead to surfaces displaying scaling laws that match those of analytical solutions (Schmittbuhl et al., 2004; Rolland et al., 2012), if the normal stress difference is taken as the main driving force. These two approaches where the main driving force is related either to gradient in normal stress or to the gradient in free energy are still debated, even among the authors of the present review. More experimental studies are needed to isolate the effect of each as a driving force.

*4.1.1 Discrete Element Models: dissolving grains.*

To treat the actual dissolution of the interface, several approaches can be taken. Koehn et al. (2007), and Makedonska et al. (2010) treat the solid using a lattice model (i.e. Discrete Element Method, DEM) that either directly uses particles, or represents particles with linear elastic springs (Aharonov and Katsman, 2009) (Fig. 18). The interface between particles can represent a grain boundary that may dissolve. In the spring model, since the dissolution is happening at an interface, and perpendicular to it, the interface movement has to be translated to the lattice spring representation. The DEM and spring methods are mathematically equivalent, since they use the same equations to calculate stress and dissolution. For both methods, one can a) dissolve particles/springs as a whole; or b) dissolve (or shorten) particles/springs partly. Figure 18a shows a DEM model, where single particles are entirely removed from the lattice and the initial surface of dissolution is predefined. Dissolution occurs in small steps, assuming that over large time steps dissolution becomes continuous. Figure 18b shows a spring model, where the initial surface of dissolution is not predefined and the dissolution is mimicked by changing the equilibrium length of stressed springs. This model resolves much smaller changes in dissolution and is useful for studying the effect of the various driving terms on the localization process.

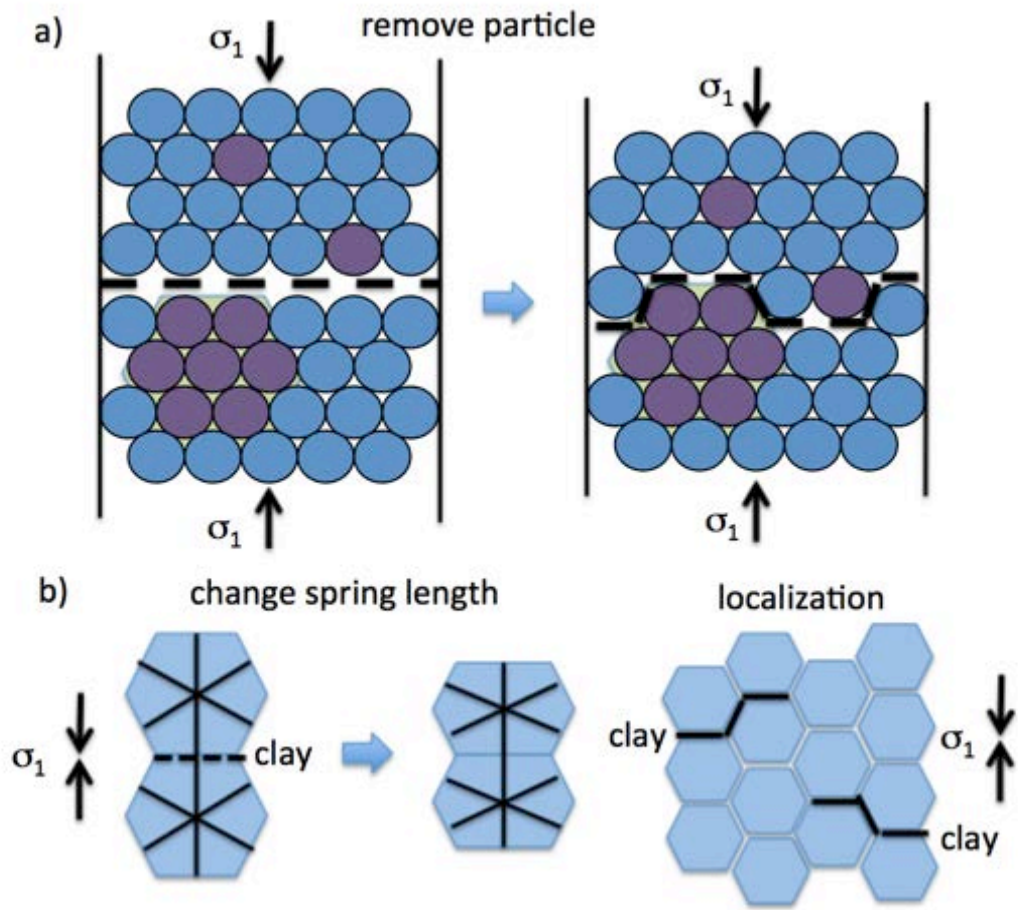

Figure 18: Model setups. a) Discrete Element Method (DEM) assumes dissolution occurs by removing particles from the lattice, according to their stress and energy. Such a model with a predefined initially flat surface, was used to study stylolite roughening. Dissolution kinetics includes a random factor, that can be distributed independently for each grain from a continuous distribution, or distributed over groups of grains from a bimodal distribution, with a slow value on purple grains, and a large one on blue grains, to study the effect of the type of disorder on the model. b) Spring network model changes the length of springs in the direction of the dissolution, using rates determined by stress and other factors.

*4.1.2 Linearized theories for roughening interfaces: analytical and numerical approaches.*

The stability of a dissolving stylolitic interface, experiencing an overall applied stress, and the stress perturbation around a stylolite due to elastic interactions was considered analytically in Gal et al. (1998), Misbah et al. (2004), Schmittbuhl et al. (2004), Bonnetier et al. (2009), and Rolland et al. (2012). These works computed the stress field from the elastostatic Green function (Landau and Lifshitz, 1986) or using an Airy function, assuming a local absence of shear along the elastic rock / soft insoluble material boundaries. The stylolite was thus approximated mechanically as an interface sustaining normal stress, without tangential stress. This state gives rise to a force perturbation related to the surface roughness, i.e. arising from the slope of the stylolite with respect to the principal stress directions, and function of the scalar values of the invariants of the far field stress (shear stress and pressure).

Rolland et al. (2012) and Schmittbuhl et al. (2004) showed that assuming that the stylolite is close to a flat interface allows the resulting stress perturbation in the elastic solid, and the associated increase of the chemical potential for the solid (i.e. the elastic part of the Helmholtz free energy, (Eq. 12)), to be expressed. Additionally, a surface energy term, proportional to the local curvature of the stylolite, is considered in the Helmholtz free energy term. The evolution of the stylolite height, as function of position and time (for a simplified two-dimensional model), reduces to:

$$\partial_t h(x,t) = v_0 + \alpha\gamma\, \partial_{xx} h(x,t) - \alpha \left(\frac{\beta P \sigma_s}{E}\right) \int_{-\infty}^{\infty} \frac{\partial_y h(y,t)}{x-y} dy + \eta(x, h(x,t)), \qquad (14)$$

where the first term on the right hand side represents an average dissolution speed, depending on the fluid composition in the stylolite, the second term is proportional to the surface energy, $\gamma$, and the third term is proportional to the elastic energy density change due to the surface shape perturbations with $E, P, \sigma_s$ the Young's modulus, the far field pressure and the shear stress, respectively. The last term, $\eta$, accounts for a quenched disorder, i.e. heterogeneities frozen in the rock, arising from the spatial variability of the material properties composing the grains. $\alpha, \beta$ are constant parameters related to the dissolution kinetics and the Poisson's coefficient of the material (Rolland et al., 2012). This time-dependent equation can be solved using a finite differences scheme in real space. The elastic kernel in the integral can also be expressed in the Fourier space and integrated numerically. This approach enables the study of the progressive roughening of the interface, that takes place due to presence of the quenched "noise", whereas a stabilization towards flattening occur due to surface tension at small scales, and elastic forces at large scales (Rolland et al., 2012).

This formulation in Eq. 14 can also be used directly for its relation to other known systems. Indeed, the spatial Fourier components of the height in these equations, depending on their wavelength, are sensitive to different terms, as summarized in Eqs. (15,16).

At large scales, the elastic forces dominate, and this equation reduces to:

$$\partial_t h(x,t) = v_0 - \alpha \left(\frac{\beta P \sigma_s}{E}\right) \int_{-\infty}^{\infty} \frac{\partial_y h(y,t)}{x-y} dy + \eta(x, h(x,t)). \qquad (15)$$

This formulation is similar to the dynamic modeling of the propagation of subcritical crack fronts in mode I, which leads to a scaling behavior of h at fixed time displaying a self-affine character with Hurst exponent around H=0.5 (Tanguy et al. 1998, Santucci et al. 2010, Tallakstad et al. 2011).

At small scales, on contrary, the surface tension dominates and the Fourier components are ruled by the Edwards-Wilkinson equation with quenched disorder (i.e. with random terms present at fixed positions, that do not evolve in time) (Roux and Hansen 1994),

$$\partial_t h(x,t) = v_0 + \alpha \gamma \, \partial_{xx} h(x,t) + \eta(x, h(x,t)) \qquad (16)$$

These two generic systems, corresponding to Eqs. (15,16), have been studied in great details in the statistical physics community, to model fracture propagation in disordered media, and surface growth or dissolution in disordered media. This was done using a variety of methods, and the rough surfaces emerging are known to follow scaling laws with known scaling law exponents in space and time. The length scale which separates these two behaviors, also called cross-over length, is the ratio of the prefactors of the two terms respectively present in Eqs. (15,16),

$$L^* = \left(\frac{\gamma E}{\beta P \sigma_s}\right), \tag{17}$$

where $\gamma, E, P$ and $\sigma_s$ are respectively the surface tension, the Young modulus, the far field pressure and shear stress, and $\beta$ is a geometric constant of order unity. Because this length scale depends on stress, it can be used to constrain the stress state and depth during stylolite formation (Ebner et al., 2009b, Rolland et al., 2012).

To study how roughness is controlled by the heterogeneities in the system, discrete elements models can also be used. The simplest case is to include these heterogeneities in the rate constant *(k<sub>i</sub>)* in Eq. (13) so that in the discrete elements numerical model shown in Figure 18a, different particles dissolve at different rates. The slowest dissolving particles or clusters of particles will then pin the interface and lead to surface roughening (Fig. 19). Surface and elastic energy will work against the pinning and tend to keep the surface flat. This "smoothing behavior" is facilitated when the pinning particles are dissolved if they meet another pinning particle across of the interface.

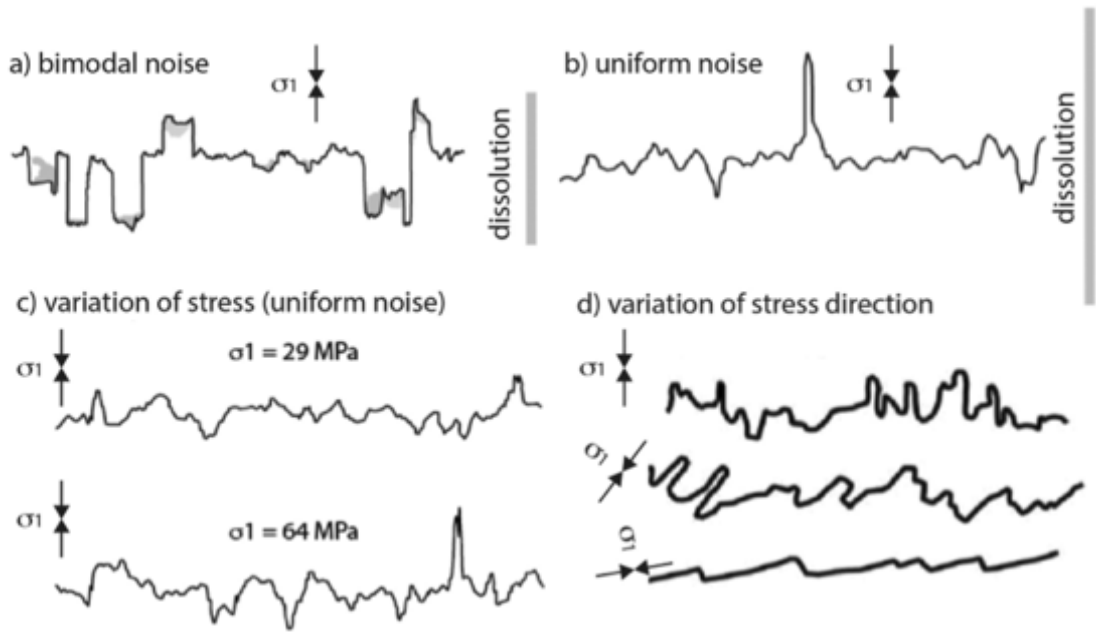

Figure 19. Numerical models of different stylolite shapes resulting from variable conditions. The total amount of dissolution accommodated is given by the vertical gray bars. a) Model of stylolite roughness with two different grain sizes (large grains are gray clusters here, and small grains correspond to the thickness of the narrowest teeth). b) Model of stylolite roughness with a uniform grain size, corresponding to the thickness of the narrowest teeth. c)

Two models of stylolites that grew under different maximum compressive stress. d) Three models of teeth growth where the initial interface is horizontal but the maximum compression has different orientations, non-normal to this initial interface. Adapted from Koehn et al. (2007, 2012).

The roughening process does develop characteristic patterns depending mainly on the direction of stress and the distribution of heterogeneity in the host rock (noise). Koehn et al. (2012) showed that a bimodal noise (i.e. dissolution constants only adopting two values) can produce large teeth whereas a uniform noise (i.e. dissolution constants fluctuating in a whole range between two extreme values) tends to produce a rough surface with some spikes (Fig. 19c). If a bimodal noise with large grains or fossils produces pronounced teeth, they can be used to estimate the amount of dissolved material at the stylolite (Fig. 19a: the amount of dissolution, corresponding to the vertical gray bar, coincides with the largest teeth size). When the noise is uniform, the length of a spike is not really representative of the dissolution at the interface, because the pinning particles at the corner of teeth are more quickly dissolved, and these teeth decrease in amplitude and become smaller spikes, so that the dissolution may be underestimated (Fig. 19b: the amount of dissolution, indicated by the vertical gray bar, is significantly larger than the size of the spikes). In addition, both the amount and direction of stress on the stylolite influences the developing roughness. Although the effect of the stress magnitude cannot really be discerned by the human eye, even though the stylolite with greater stress seems a bit more spiky (Fig. 19c), it can be quantified via a roughness analysis. Finally, the direction of the main compressive stress is always parallel to the sides of stylolite teeth, so that the teeth are inclined if dissolution surfaces are not normal to the maximum compression axis (Fig. 19d).

*4.2 Different aspects of stylolite formation*

*4.2.1 Models of Initial localization and lateral propagation*

The development of the initial localization that generates a stylolite is debated. The simplest case is to assume that the interface is present in the original rock, for example material differences across bedding planes for sedimentary stylolites or veins or joints in the case of tectonic stylolites. Evidence for preexisting interfaces acting as loci for stylolitization includes observations that sedimentary stylolites often localize along stratifications and laminations (Stockdale, 1922) and observations of >1 km long stylolites formed along bedding planes (Ben-Itzhak et al., 2012). In addition, stylolites often form along pre-existing joints and faults (Stockdale, 1922; Geiser and Sansone, 1981). However, since some stylolites form complex networks, and tectonic stylolites may also form without an underlying heterogeneity (at scales larger than the granular one), we assume that some stylolitic localization develops without the presence of pre-existing transgranular surface, during the initial dissolution process. As reviewed by Engelder and Marshak (1985) *"morphology and distribution are controlled by several parameters including: rock composition, structural position and deformational environment"*. At a given location, different lithologies typically develop different structures. For example, pure quartzites or pure limestones deform predominantly by intracrystalline or brittle mechanisms, whereas impure (clay-rich) sandstones and limestones develop cleavage (e.g. Heald, 1956; Weyl, 1959; DeBoer, 1977; Wanless, 1979; Morris, 1981; Marshak & Engelder ,1985). Heald (1956) concluded that "*pressure solution may occur without clay, but if other conditions are favorable, clay accelerates the process*." It was also observed that clay-matrix content in sandstones and limestones controls domain spacing, because spacing decreases as clay content increases (e.g. Wanless, 1979).

Four main processes have been proposed for the localization and/or propagation of stylolite tips: i) stress – which aids in localization of pressure solution by concentrations at the tips of dissolution surfaces, so that stylolites can grow at tips, similar to the anti-crack concept of Fletcher and Pollard (1981), and the dislocation model of Katsman et al. (2006a,b); ii) strain-energy controlled pressure solution that occurs on free-faces of grains increases porosity, increasing stress and promoting further pressure solution in a high-porosity stylolite-bearing layer (Marino et al,. 1983); iii) dissolution enhancement due to clay content leading to localization (Aharonov and Katsman, 2010); iv) dissolution enhancement via other rock heterogeneities such as cracks allowing enhanced transport-controlled dissolution (Ben-Itzhak et al., 2014), or a layer of small grain-size locally increasing free energy (Rustichelli et al., 2015).

All stylolites occur by localized dissolution, yet lateral propagation of stylolite tips is observed mostly for tectonic stylolites (e.g., Figure 3 in Rispoli, 1981). Sedimentary stylolites are mostly found in networks - either very long and parallel stylolites (Fig. 3) or as interconnected stylolites with or without fractures (Fig. 7 a&b). For many sedimentary stylolites, free-standing tips are seldom observed (Ben-Itzhak et al., 2014), and there is little or no evidence exists for lateral propagation during their formation. An example where lateral propagation seems to have occurred, and where the stylolites were sensitive to the stress perturbation due to the presence of other stylolites, is illustrated by the two interacting terminations of stylolites in Figure 7c. The teeth amplitude diminishes toward the terminations, and the two stylolites ends are deflected around each other.

Lateral propagation of stylolites or solution seams can depend on composition and stress concentration: (i) initiation of dissolution seam located in a zone of maximum stress that, due to soluble species depletion, becomes a relatively weak zone (e.g. Gratier et al, 2015); (ii) lateral propagation perpendicular to the maximum compressive stress due to stress concentration at the tip of the relatively weak seam (Cosgrove, 1976; Gratier, 1987). The pioneering anti-crack model was proposed by Cosgrove (1976) and Fletcher and Pollard (1981) to explain lateral stylolite tip propagation via stress feedback. Their idea was that dissolution produces an elliptical hole, or in Cosgrove's model, a weak domain filled with clay residue, with stress concentrations at its tips. These stress concentrations drive further dissolution leading to growth of a penny-shaped plane that develops perpendicular to the main compressive stress, in a manner similar, but opposite in sign, to a mode I crack lengthening. Their model assumed that since the hole or the dissolution zone are relatively weak under stress, stress becomes increasingly elevated on flanks, which drives continued pressure solution on flanks and further concentrates the stress at the tips, allowing lateral propagation in a self–similar manner.

Katsman and Aharonov (2005) challenged this model and calculated that the closing ellipse will not have high stress on it flanks. Instead, the stress formed will be similar to an edge dislocation (and not to an anti-crack), with reduced stress on its flanks. This conceptualization creates a problem for the anti-crack model because: i) The stress concentration at the tip will be quickly smeared by dissolution unless dissolution continues on flanks, ii) without stress elevation at the flanks, no stress-driver exists for continued pressure solution on flanks. To overcome these issues, and also to incorporate the input from field observations for the enhancing role of clay, Katsman et al. (2006a, 2006b) and Aharonov and Katsman (2009) proposed a new model, where localization is aided by clay minerals present at grain boundaries, enhancing the dissolution process. Enhanced stresses at tips control lateral propagation of stylolite, while clay-enhanced dissolution at stylolite flanks, where stress is low, thickens them. Such thickening, with a resulting increased thickness growing with increasing isolated stylolites length, has been measured, e.g. (Benedicto and Schultz, 2010); Nenna and Aydin, 2012). Therefore, dissolution progressively concentrates clays, and as clay abundance increases, it enhances the dissolution process, leading to localization along seams and development of networks. It is well known that stylolites do contain clays (e.g. Nenna and Aydin, 2012), however the exact nature of the enhancement of pressure solution because of clays, and its role in creation of stylolite networks, is still a topic for further research.

After stylolite nucleation at point or linear heterogeneities, followed by lateral growth and thickening (in case of growth from a point heterogeneity), or followed by only thickening (in case of stylolites growing from a linear heterogeneity such as bedding plane), stylolites may cross each other when the rock matrix between the planes is entirely dissolved. This interaction results in the merging of portions of stylolites, also referred to as an anastomosing geometry. Sketches of this stages of development for stylolite networks including strain distributions have been developed (Fueten et al., 2002; Nenna & Aydin, 2012) (Fig. 7, video in Supplementary Materials). A cartoon of this process is illustrated in Fig 22c.

*4.2.2 Scaling of stylolite surface in space and time: analytical and numerical results*

Starting dissolution from an initially flat interface, and distributing the quenched disorder associated to material properties randomly to the grain population in the surrounding half-spaces for modelling, without any correlations between these values, one can integrate directly by finite differences the equation (14) (Rolland et al., 2012) to simulate stylolite intiation and growth as a set of self-affine surfaces. These surfaces have Hurst exponents $H=1$ to 1.2 at small scales, and $H=0.35$ to 0.4 at scales above the cross-over length scale of Eq. (17). This is measurable via the Fourier power spectrum of the stylolite surface elevation with respect to its average plane: in accordance with a self-affine property, these two scaling regimes correspond to Fourier mode amplitude being power-laws of the wavelength, i.e. they are straight lines in a bi-logarithmic plot, with slopes corresponding to $1+2H$, where $H$ is the Hurst exponent (Fig. 20).

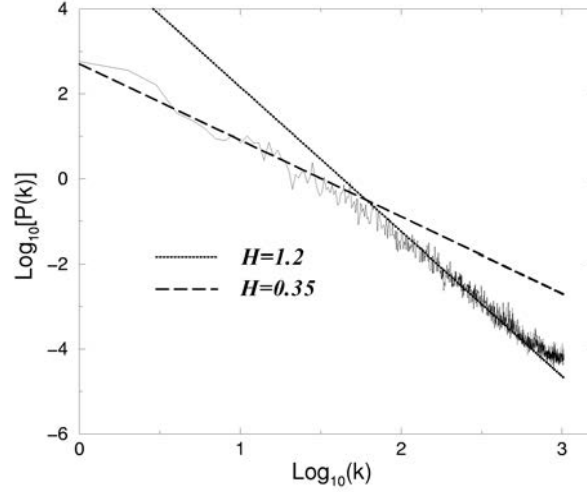

Figure 20: Fourier power spectrum of the developed interfaces in bi-logarithmic representation. The length unit is 10 micrometers. The two scaling relationships observed correspond to power laws of exponents -1-2H, with Hurst exponents respectively at H=0.35 and 1.2. The cross-over is at $k^* = 2\pi/L^*$, where $L^*$ is defined in Eq. (17). After Rolland (2012).

These scaling exponents are the same as for elastic line growth models, Eq. (7) in Tanguy et al., 1998, and the Edwards-Wilkinson model with quenched noise (Roux and Hansen, 1994). In both models, the scaling exponents in time and in space are related, i.e. the dynamics of growth follows what is called in statistical physics a Family-Vicsek scaling (Family and Vicsek, 1985): first, a correlation length $\xi$ for the random component of the height deviation from an average plane grows as:

$$\xi \sim t^{1/z} \tag{18}$$

with $z$ being the so-called dynamic exponent. Next, calling $w(l, t)$ the width measured at scale $l$ and time $t$ after the beginning of the roughening, where the methods for measuring $w$ are defined in Section 2.4, $w(l, t)$ follows a Family-Vicsek scaling of the form:

$$w(l, t) \sim l^H f\left(\frac{l}{\xi}\right), \tag{19}$$

where H is the Hurst exponent. This scaling relationship also assumes that the width of the stylolite is independent of the correlation length for scales below it, and is independent of $l$ above the correlation length, so that:

$$f(x) \propto x^{-H}, (x \gg 1) \tag{20}$$

$$f(x) \sim constant, (x \ll 1) \tag{21}$$

The corresponding dynamic exponents are reported (Schmittbuhl et al., 2004) as $z =1$ to 1.2 for the elastic roughening model of Eq. (14) in Tanguy et al. (1998), and $z =0.8$ (Roux et al., 1994) for the small-scale surface energy dominated regime. In general Family-Vicsek scaling law relating the temporal and spatial scaling is commonly observed for roughening lines or interfaces, as propagating fracture fronts (Tallakstad et al., 2011), dipolar chains in fluids (Toussaint et al., 2004, 2006) or sedimentation fronts (Vinningland et al., 2012).

The stylolite roughening can also be modeled in using DEM, and the space and time scaling demonstrated in the previous perturbative modelling can also be explored in this type of model. Using discrete element models, both the time and length scales are shown in Figure 21a-b where the lefthand side shows a short stylolite growing in the surface-energy-dominated regime and the righthand side is a 100 times longer stylolite, growing in the elastic energy dominated regime. The small-scale stylolite is relatively smooth, grows slowly in time and stops growing in amplitude after numerical time step 100 (Figure 21a). In contrast, the large-scale stylolite is rough with pronounced teeth, grows faster in time and does not stop growing. Figures 21c-d illustrate the space and time scaling, where the height-height correlation functions displays a linear scaling in plots of log wavelength versus log of correlation function amplitude, showing that the growth is self-affine. The slope of the lines in the log-log plot yields the roughness exponent of the interface. A roughness exponent of 0.8, close to 1.0, is characteristic of surface-energy-dominated growth (Eq. 16), whereas a roughness exponent between 0.3 and 0.5 is characteristic of elastic-energy-dominated growth (Eq. 15). Both regimes can be found in a single stylolite as a function of scale and the operative control for growth (Fig. 21c). The stylolite surfaces can contain both geometries for surfaces where both regimes are operative at their respective scales.

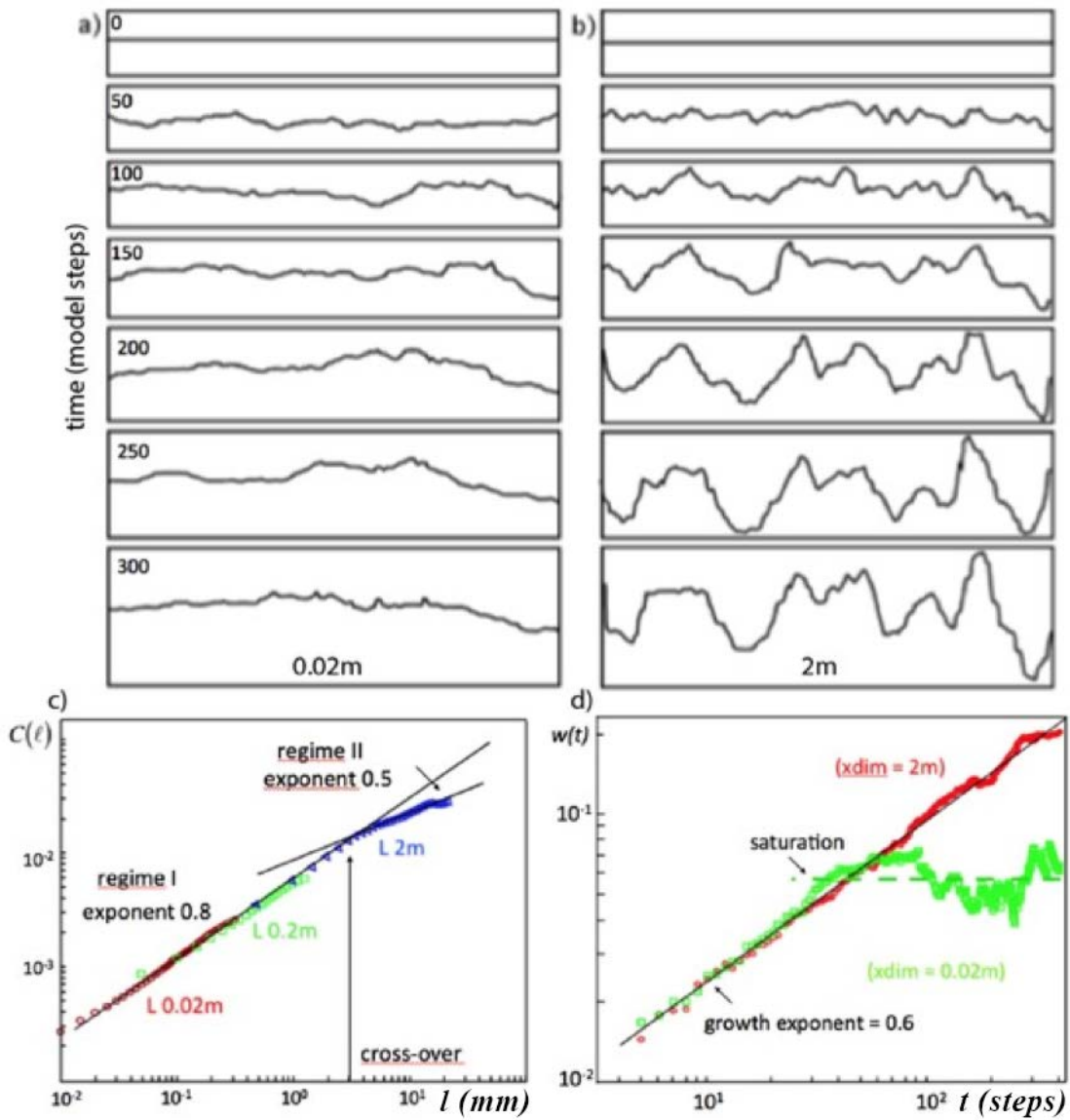

Figure 21: Discrete element modelling of stylolite roughness development at different scales and growth in space and time. a) Growth at the small scale (dimension 0.02m) showing a surface energy dominated stylolite with H=0.8, close to 1. b) Growth with pronounced teeth when elastic energy dominates, with H=0.5. c) Model simulation where both controls are operative as a function of scale so that both geometries are present and are separated by a cross-over scale: Scaling law in space illustrated using a correlation function as function of space lag, with two power laws (slopes in logarithmic representation) and a cross-over length-scale separating these two. d) Scaling of stylolite growth as a function of regime as function of time, showing that the small stylolite roughness saturates in amplitude, whereas the long one roughens at later time.

The growth of the roughness of interfaces in time is revealed by considering the mean width of the interface, which gives an average roughness amplitude for each time step. A log-log plot of the mean width versus time shows a linear trend consistent with a characteristic growth exponent (H=0.6 in Fig. 21d). If the system size is fixed by the number of particles along the x-axis, the growth can saturate when the roughness amplitude reaches a maximum value. This saturation is a finite-size effect, since larger system sizes lead to larger saturation roughness amplitude, and presumably infinite systems would present ever-growing amplitude. This outcome is clearly seen in the surface-energy-dominated stylolite that only grows in the first 100 numerical time steps and then the roughness amplitude does not increase any more (Fig 21d). Model results indicate that saturation occurs more rapidly for shorter stylolites involving fewer grains.

Natural stylolites may not have a system size, especially sedimentary stylolites that may be very long. However, tectonic stylolites growing within restricted bedding planes show a roughness saturation. In general, scaling of the roughness amplitude with space shows a large Hurst (roughness) exponent on the small scale and a lsmaller Hurst exponent at larger scales, followed by a saturation of the roughness amplitude depending on how long the stylolite grew, i.e. how much material was dissolved. Similar scaling can be seen in time. Ebner et al. (2009b) have shown that surface-energy-dominated growth and elastic-energy-dominated growth may have different scaling exponents in time, followed by a saturation of the growth depending on time or amount of dissolution and system size. In addition, Ebner et al. (2009a) showed in details how the scaling behavior in space and time is influenced by the noise in the system. For example, Koehn et al. (2012) demonstrated that stylolites in a host rock with a bimodal grain size distribution including large pinning particles record their full growth history for the stylolitic roughness and thus have a growth exponent of 1.0 as long as the large pinning grains or fossils are not completely dissolved.

*4.2.3 Models investigating stylolite distributions or networks*

***Formation of long parallel stylolites***: Merino et al. (1983) proposed a model for the formation of simultaneous parallel stylolites, occuring spontaneously when pressure solution is controlled by strain energy and occurs on free-faces of grains. In the model, free-face pressure solution, as opposed to thin-film pressure solution, preferentially dissolves high strain and stress regions of grain contacts, reducing grain-grain contact sizes. The reduced area of contact between grains increases contact stress, which in turn increases the rate of pressure solution and reduces contact size even further, although experiments do not show this instability and instead suggest that a steady-state grain size can be reached (Karcz et al., 2008). Diffusion of solute and re-precipitation reduce porosity and decrease stress in regions adjacent to stylolites, thus localizing the dissolution even further on the high porosity proto-stylolites. Besides, Merino et al. (1983) model predicts the formation of infinitely long, evenly spaced and very high porosity stylolites with highly cemented regions in between. Railsback (1998) however performed statistical analysis of stylolite spacing and suggested that stylolites observed in the field are spaced at a random distance from each other.

Another mechanism has also been proposed to explain how series of long parallel stylolites interact with their surroundings. Based on field observations in sandstones from the North Sea that related enhanced pressure solution and stylolites to clays, Wangen (1998) suggested that a stylolite is a clay-lined surface with a higher solubility with respect to quartz than its surrounding. Considering a second factor, for carbonates, Rustichelli et al. (2015) showed that a layer with different grain sizes caused bedding-parallel stylolites to form. In particular, small grain sizes are expected to enhance solubility. These two heterogeneities trigger a solubility jump that can cause the solute to be super-saturated with respect to outside localities. Thus, the dissolved minerals from the stylolite become a source of cement to the surrounding region. This conceptual framework enabled analyses of the controlling parameters for stylolite dissolution and surrounding cementation. Further, predictions about of rates of stylolite dissolution, porosity loss along with the spatial distribution of new cement, can be made (Angheluta et al., 2012 (Fig. 17, Section 3.3)). Field measurements by Ben-Itzhak et al. (2012) provide support that sedimentary parallel stylolites form on initial bedding planes, that continuously roughen via dissolution, via the rougnening models (Koehn et al., 2007; Ebner et al., 2009a, Rolland et al., 2012). Statistical analysis of measurements for stylolite roughness, and mass balance calculations, further suggest that dissolution continues until all macro-porosity in surrounding area was filled with cement. In contrast to macroscopic porosity, models and field observations (Emmanuel and Ague 2009; Emmanuel et al. 2010) suggest that nano-scale porosity may remain fixed even as the rest of the pore space clogs up, due to the high curvature of small-scale pores which increases their solubility.

*Formation of anastamosing networks*: These geometries are the least studied of all types of stylolites. Fueten and Robin (1992) and Fueten et al. (2002) modeled the formation of connected networks of solution seams from an initial distribution of clay partings. The clay partings were modeled as having low viscosity, and connected via stress-induced dissolution at their tips due to their softness, into networks of solution seams. Drawing on prior work, Ben-Itzhak et al. (2014) presented three concepts for the formation of connected networks (Fig. 22): (a) connection of isolated stylolites via dissolution at tips (Fig. 22a); (b) stylolites connect by mode I cracks (enhancing fluid flow that enhance stylolite formation) or shear fractures (forming at the stylolite tips due to the stress perturbation they induce) (Fig. 22a-b); or (c) via "cannibalism" of long parallel stylolites.

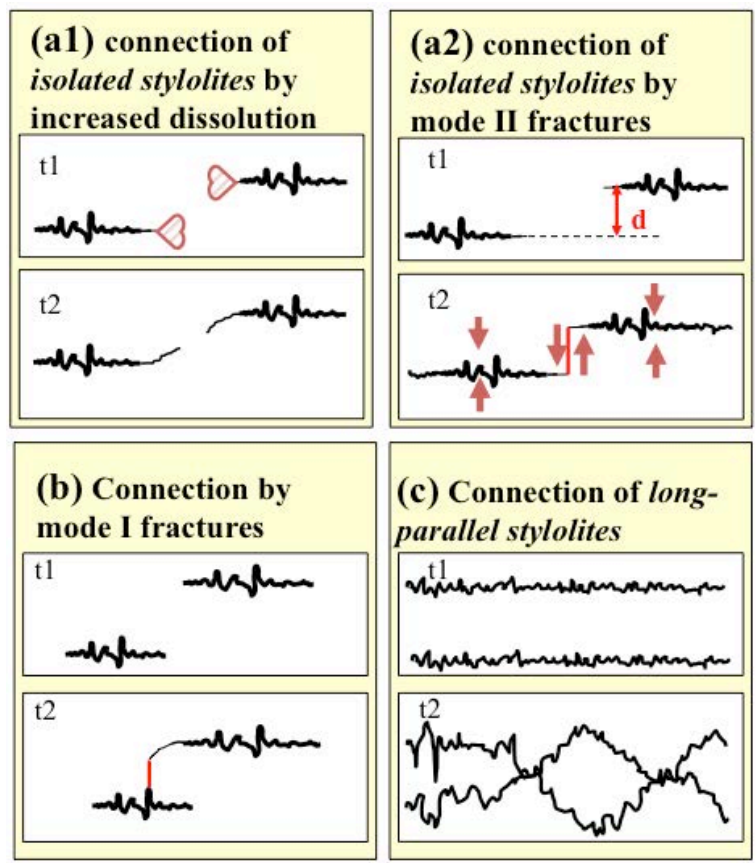

Figure 22: Cartoon showing several possible mechanisms for the formation of interconnected stylolite networks between times t1 and t2. a) Connection of isolated stylolites. Increased normal stress in heart-shaped-regions enhances dissolution and may change propagation direction (a1), while increased shear strain may cause fracturing (a2). b) Connection of stylolites via mode I fractures emanating from stylolite teeth. Tips of an adjacent stylolite are seen to curve towards veins. This scenario fits with stylolite localizing via high dissolution rates. Rapid dissolution in diffusion limited systems can be induced by rapid fluid pathways. c) Connection of long-parallel stylolites by their dissolution and roughening -"cannibalism" of closely spaced long-parallel stylolites. Adapted from Ben-Itzhak et al. (2014).

## 5. Laboratory stylolites: the experimental challenge

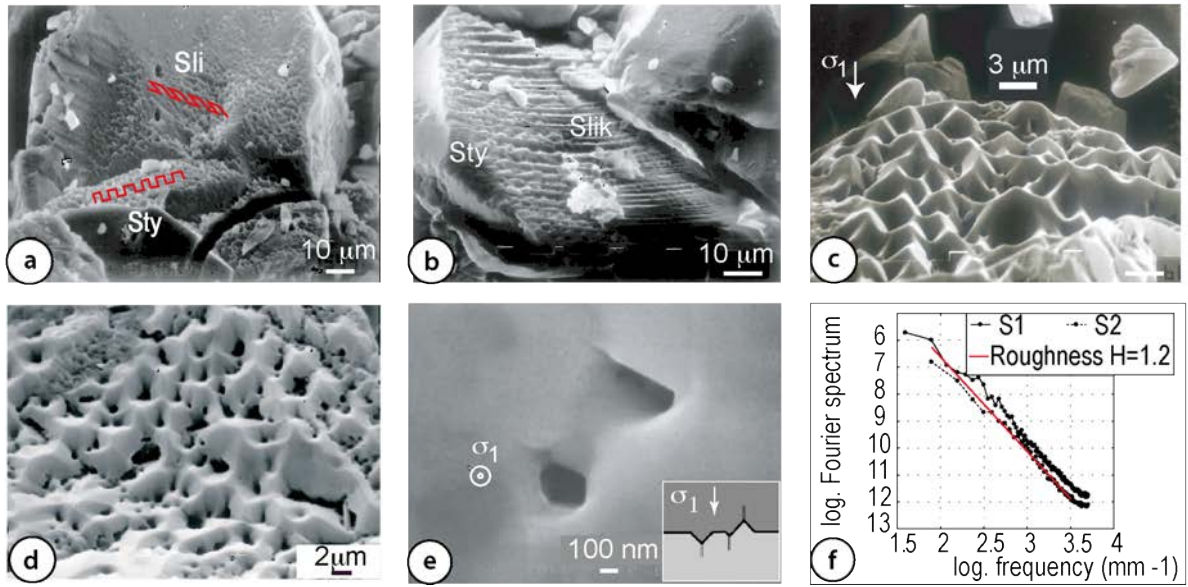

Figure 23: Experimental microstylolites on quartz grains: (a) microstylolites (Sty) and slickolites (Sli) with peaks and spikes parallel to the maximum compressive stress; (b) microstylolites changing to solution striation (Slik) due to grain rotation; (c) microstylolite peaks in quartz; (d) evidence of dissolution pits developed in front of each microstylolite peak; (e): axial view of dissolution pits along a dislocation at the bottom of a microstylolitic peak and cross sectional view in inset; (f) Spatial wavelength analysis of the microstylolite surface by Fourier transform: the surface are self-affine with a linear slope in a log-log plot for a Hurst roughness exponent of 1.2. Adapted from Gratier et al. (2005).

It is very difficult to reproduce stylolites in the laboratory. Despite many attempts, few lab-generated stylolites have been created and they are typically only microstylolites present at only the scale of one to a few grains (Gratier et al., 2005). Microstylolites were created at the grain scale on various minerals such as calcite (Hellmann et al., 2002; Zhang et al., 2010) and quartz (Gratier and Guiguet, 1986; Gratier et al., 2005; Niemeijer et al., 2002, Gratier et al., 2009). They are geometrically like stylolites as they have both the dissolution features and have undulated shapes. Unlike other studies where microstylolites were only seen in 2D, the studies of Gratier et al. (2005) examined experimental stylolites in 3D as to analyze their 3D geometry while comparing them with natural stylolites to identify similar characteristics (Fig. 23). Experiments were performed on quartz with constant compressive stress with reactive fluid (pure water or 0.1M NaOH saturated solution) in long duration experiments (43-51 days, exceeding the duration of classical chemistry experiments in fluid phase). Experimental stylolites (Fig. 23a) show a continuous transition between (i) stylolite surfaces orthogonal to the displacement direction; (ii) slickolite surfaces orthogonal to the displacement direction; and (iii) solution striations parallel to displacement. This continuity means that all stylolites peaks are parallel to the imposed maximum compressive stress in a co-axial regime, as would be expected. However, this experiment also showed that solution striations may develop oblique to the maximum compressive during grain rotation (Fig. 23b). Moreover, image analysis of the stylolitic surface (Fig. 23d,f) show a fractal self affine roughness. The Hurst exponent H = 1.2 indicates the dominant role of the surface energy at this scale (Renard et al., 2004; Schmittbuhl et al., 2004). Such experimental stylolites also show that, at small scale, the quenched material disorder required in stylolitic modelling (Renard et al., 2004; Schmittbuhl et al., 2004) could be due to the presence of dislocations given that microstylolitic peaks are rooted on dissolution pits where a dislocation is present at the stylolite surface (Gratier et al., 2005) (Fig. 25e). Other compaction experiments of quartz aggregate compaction performed at 150 °C and a tenth of MPa of pressure showed important pressure-solution occurring, but no stylolite formation was reported (He et al., 2003).

Why is it so difficult to make macrostylolites experimentally, several grains large in the direction transverse to the stylolites? At the grain scale, microstylolite wavelengths of no more than some micrometers are obtained by experiments after weeks or months of duration, in the best possible conditions: relatively high differential stress (within the limit of grain fracturing), and enhanced mineral solubilities. Developing stylolites at multigrain or rock scale would mean generating wavelengths greater than grain size. Using limestone or sandstone with grain sizes of several hundred micrometers requires generating stylolite wavelengths of several millimeters, and that is on the order of magnitude of thousand times the size of the microstylolites that have been experimentally developed to date. If the relationship between growth rate and time is linear, which seems plausible, creating experimental macrostylolites would require two thousand months, or more than hundred years in natural rocks. Such growth rates for stylolites are supported by model simulations (Angheluta et al., 2012). Consequently, to gain an understanding of stylolite development experimentally, it is necessary to find proxy materials that will yield appropriate geometries for similar deformation behaviors and do so in much shorter timeframes. Recently, a useful proxy, a mixture of gypsum plaster and illite (Gratier et al., 2015) has opened the possibility of reproducing stylolites experimentally in analogous material such as gypsum plaster or halite aggregate mixed with clay minerals.

What would be the best conditions to obtain stylolites in the laboratory? Considering model results, one may need very dense quenched noise co-located with compositional and grain-size heterogeneities at the grain scale. So a mixing of soluble and insoluble species would be favorable. However, since stylolite seems to be restricted to host rocks with initial high contents of soluble species (Fig. 15), the amount of insoluble species must stay very low. Heterogeneities in the soluble minerals would also help. Also, because stylolite wavelength must be larger than the grain size, one can choose grain size as small as possible, knowing however that for very small grain size the effect of the surface energy may prohibit stylolite development. The stress driving force cannot exceed that used for the development of microstylolites since it was already near the yield strength of the grains and the solubility cannot be increased more than for microstylolites except if high solubility minerals are used, such as salts. One must probably start from a pre-existing flat surface, so as to reproduce sedimentary stylolites, rather than tectonic stylolites. However, even in the best conditions with a grain size of 10 micrometers and stylolite wavelengths slightly greater than 100 micrometers, it would take roughly hundred times longer to make macrostylolites than microstylolites. So, experiments would still run several tens of years, and proxies mineral as salts at room temperature could take years. To our knowledge, such challenging experiments remain to be done.

**6. Stylolites – what can they be used for?**

*6.1 Strain markers: How to quantify the dissolution associated with stylolites from mass balance and vein displacement?*

Geometric methods may be used to determine the amount of dissolution with stylolites. The first approach is to measure the height of the stylolite peaks. However, such a measurement only gives a minimum value of the dissolved layer, assuming that all roughness was generated during stylolite formation. Moreover, when insoluble species accumulate in the stylolite seams, the stylolite shape can be smoothed. Another method is to measure the apparent displacement of veins due to the dissolution (Fig. 24). The progressive dissolution leads to an apparent lateral shift of the vein $d_s$. The dissolved width $X$ is equal to the product of the measured shift $d_s$ by the tangent of the angle $\alpha$ of the veins with the solution seam, $X = d_s \tan \alpha$.

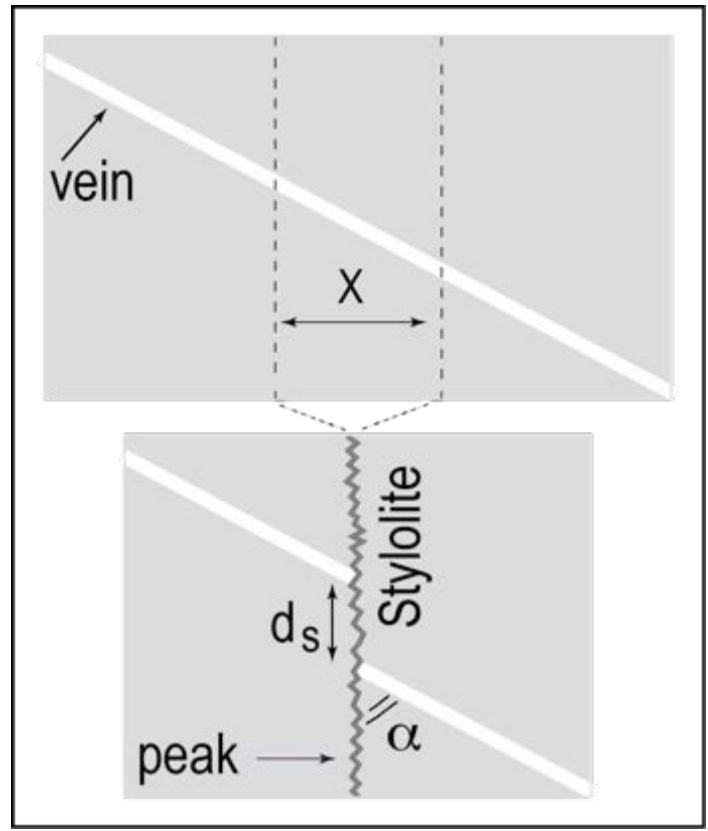

Figure 24: Use of the apparent shift of the vein $d_s$ along dissolution seams to evaluate the dissolution magnitude $X = d_s \tan \alpha$. Adapted from (Gratier et al., 2013a).

A geochemical method can also be used, based on the idea that the amount of insoluble species passively concentrated in the stylolite seams compared with the initial amount of the same insoluble species in the undeformed rock gives the amount of dissolution, assuming an initial homogeneous distribution of the species. Initial spatial heterogeneities in insoluble species will create error for this method. Schematically, the relative mass balance $\Delta M/M_0$ can be calculated using the passive concentration of insoluble minerals due to dissolution in the exposed zone compared with the (relatively) protected zone:

$$\Delta M/M_0 = I_p/I_e - 1 \tag{22}$$

$I_p$ and $I_e$ being the concentration of all insoluble minerals in the protected and in the exposed zones respectively, and $M_0$ the mass of a representative elementary volume before deformation (Fig. 25). The volume change may also be calculated if the difference in density between the two parts of the rock is known (Grant, 2005; Gresens, 1967).

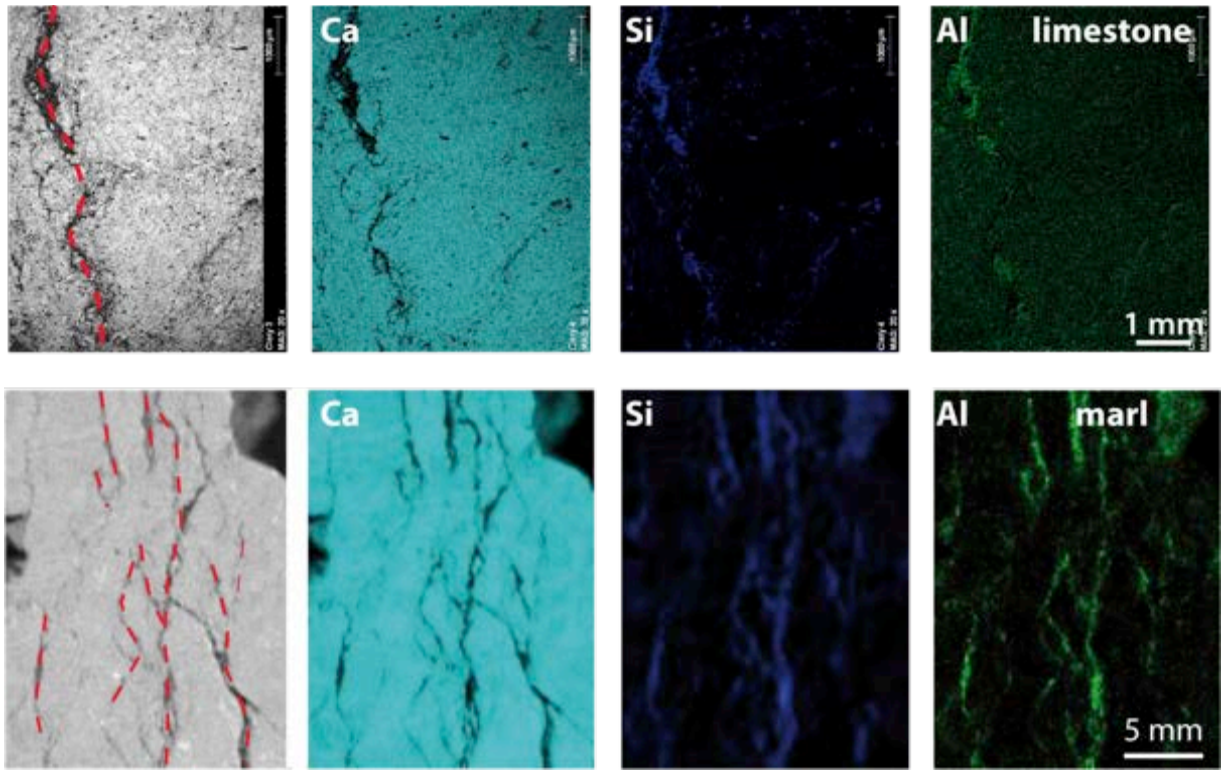

Figure 25: Element distribution from scanning electron microscopy imaging of stylolite in limestone (top) and of solution cleavage in marl (bottom). The stylolites and solution cleavage are marked in red on the left figures. In the next panels, brighter colors indicate higher content, for Al, Si and Ca elements. Dissolution of calcite explains the low Ca content in the dissolution seams (S) and is associated with passive concentration of insoluble species (Al and Si in phyllosilicates). Adapted from Gratier et al. (2013b).

*6.2 Strain markers: how to quantify the displacement across stylolites from teeth morphology*

Geologists have long used the maximum tooth amplitude in stylolites as a measure of the total amount of dissolution. However, for large amounts of dissolution, the least soluble grains may dissolve, causing this measure to to be a minimum estimate of the dissolution accommodated by the stylolite. Alternatively, it is possible to use another proxy, based on the self-affine character of the stylolite up to the correlation length, and the fact that this correlation length grows as dissolution proceeds. This proxy, developed by Ben-Itzhak et al. (2012), is based on the fact that the out-of-plane width of a stylolite follows a Family-Vicsek scaling, as detailed in Eqs. (18-20): calling $w(l,t)$ the width measured at scale $l$ and time $t$ after the beginning of the roughening, and as long as $l < \xi$, we have:

$$w(l,t) \sim \alpha l^{*H-1} \xi^H, \tag{23}$$

where $H$ is the Hurst exponent, $l^*$ a lower length scale where this scaling starts, $\alpha$ a dimensionless constant, and a correlation length $\xi$ evolving as:

$$\xi \sim t^{1/z} = \left(\frac{A}{v}\right)^{1/z}, \tag{24}$$

with $A$ is the total thickness of material dissolved by the stylolite (a length), $z$ the dynamic exponent and $v$ the average dissolution speed. As a consequence, the saturation width becomes:

$$\frac{w(\xi,t)}{l^*} \sim \alpha \left(\frac{\xi}{l^*}\right)^H = \alpha \left(\frac{A}{l^*}\right)^{H/z}, \tag{25}$$

One can thus evaluate this length $A$ from the measured values of the correlation length $\xi$, or from the saturation root-mean square width w. For stylolites in the Blanche cliff (Ben-Itzhak et al., 2012), the constants were evaluated to lead to a length scale $l^* \sim 0.5 - 3$ mm, and a prefactor $\alpha \sim 0.17$ to 0.36. Using exponents H=0.65 and H/z=0.8 (Koehn et al., 2007, Ebner et al., 2009a; Ben-Itzhak et al., 2012), the estimated amount of material dissolved $A$ is similar to values obtained using maximum teeth amplitude, and also similar to an estimate from a mass balance analysis assuming that all the dissolved material was precipitated in pore space between the stylolites.

Koehn et al. (2016) used a stylolite classification to estimate compaction at stylolites and show that the material dissolved can be as much as 30 to 40% of the initial porous host-rock. Stylolites that develop in clay layers and stylolites with pinning fossils have the best tracking capabilities, grow linearly and capture most of the actual dissolution at the stylolite.

*6.3 Stress markers: how to quantify the paleostress related to stylolite formation?*

**Paleostress orientation:** Stylolites usually develop perpendicular to the largest principal stress, and are largely used as markers of stress orientation on the field. However, when they develop on an already existing interface that was not initially normal to the direction of maximum compression, the main direction of the teeth sides coincide with this direction – leading to slickolites, which is supported by results from numerical models (Koehn et al. 2007, and Fig. 19d). These teeth are good markers for the direction of $\sigma_1$ (Figs. 6, 19d).

***Paleostress magnitude in sedimentary stylolites***: The physics of stylolite growth results from a balance between surface-energy-based stabilization mechanism and disorder at small scales, and a balance between elastic-energy-based stabilization and disorder at large scales: the two stabilizing terms in Eq. (14) dominate, for one of them, at small scale, and for the other, at large scale. Eq. (14) reduces to Eq. (15) or (16) depending on the scales considered, and these two models lead to two different scaling laws for the self-affinity of a developed stylolite (Rolland et al. 2012). The fact that these scaling laws are separated by a crossover length scale dependent on the stress tensor, from Eq. (17), $L^* = \left(\frac{\gamma E}{\beta P \sigma_s}\right)$ means that the amplitude of the paleostress, during the formation of the stylolites, leaves a marker in the geometry of these stylolites.

Measuring this change of regime between two different self-affine scaling relationships allows us to estimate $L^*$, as shown in the two examples in Fig. 26: for an open 2D stylolite in a), and for a 1D stylolite profile extracted from a borehole core sample in b).

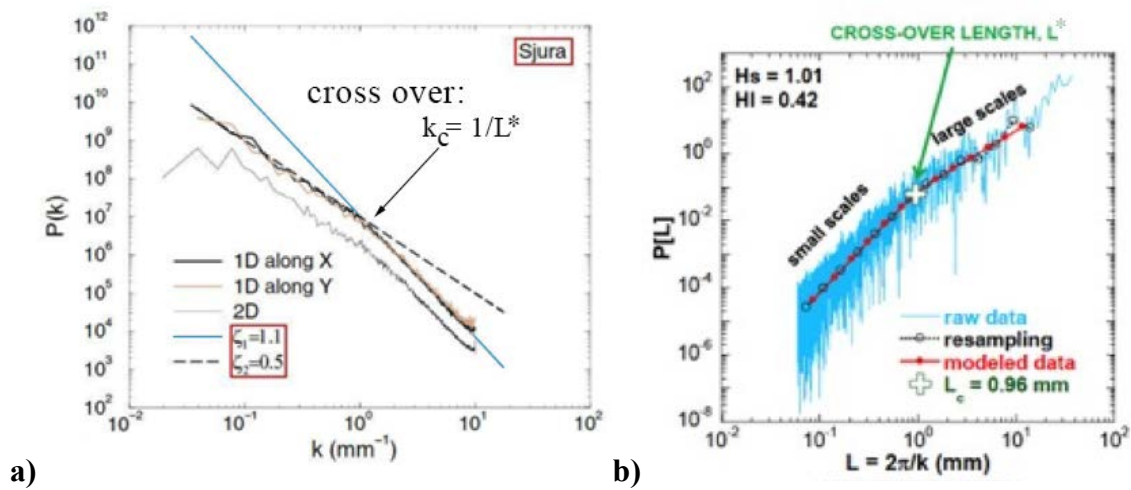

a)    b)

Figure 26: a) Fourier power spectrum for parallel profiles of the 2D roughness of an open stylolite (sample Sjura, from Fig. 1). After Schmittbuhl et al. (2004). b) Fourier power spectrum as function of the wavelength of the 1D height profile of a stylolite from the Paris Basin. After Rolland et al. (2012). Both samples show a clear crossover length scale $L^*$, corresponding respectively to 1.0 mm (a) and 0.9 mm (b).

Assuming simple vertical compaction with no horizontal movement to relate pressure and shear stress, we can use Eq. (8) to estimate the normal stress exerted on the stylolite during formation. In the case of the stylolites collected over a vertical cross-section along a cliff of several hundreds of meters height in the Cirque de Navacelle, Southern France (Ebner et al., 2009b), a clear correlation exists between the current depth in the stratigraphic column, and the formation stress as measured by the cross-over length scale $L^*$ (Fig. 27). This link between the crossover length scale and formation stress was also verified using numerical models (Koehn et al., 2012). It was used to determine the paleostress magnitude, i.e. the formation vertical stress and largest shear stress magnitudes, for series of sedimentary stylolites in the Paris Basin, ranging from 150 to 750 m depth, in the Dogger and Oxfordian Formations (Rolland et al., 2014). Figure 28 shows the vertical component of formation stress $\sigma_{zz}$ as function of depth. It allowed to conclude that the stylolites in the Dogger Formation, at z<500m, are either still active or stopped growing at larger depths than the current one, before 400 m of overburden were eroded. It also showed that most sedimentary stylolites in the Oxfordian Formation stopped growing at $\sigma_{zz}$ around 10-20 MPa, most likely because horizontal stress became lower than the vertical one at larger depths, promoting tectonic stylolites to grow instead of sedimentary ones when larger depths were reached.

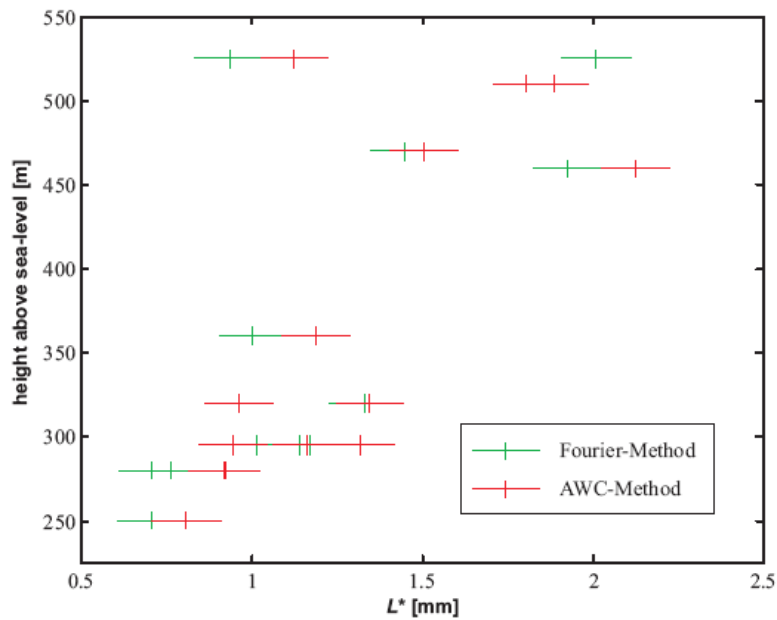

Figure 27: A crossover length scale $L^*$ determined from the power spectrum of stylolite profiles, with a Fourier Transform or an Average Wavelet Coefficients Spectrum (Ebner et al., 2009b), for sedimentary stylolites found in a large limestone formation in Southern France.

The height at which the stylolites were found at Cirque de Navacelle is shown to correlate with $L^*$, supporting the theoretical result of a link between formation stress magnitude and crossover length scale for the stylolites. When this technique is applied on a series of stylolites in different rock formations at different depths, it is possible to evaluate the formation stress of the different stylolite families.

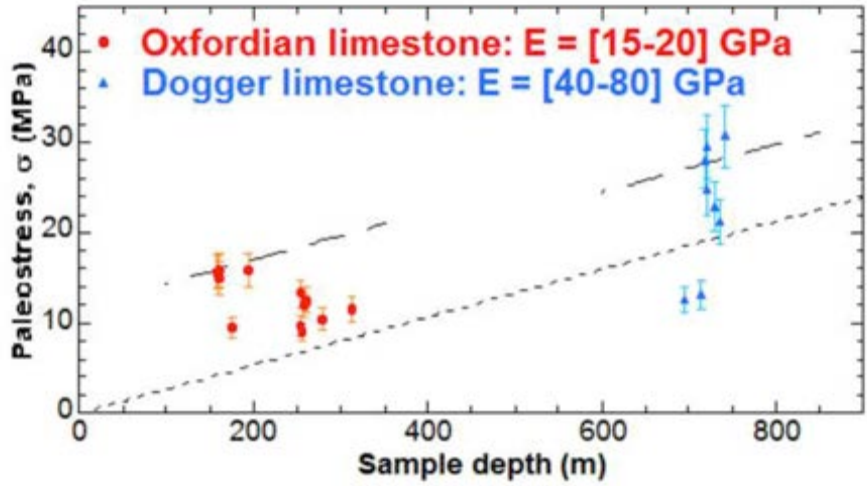

Figure 28: Formation stress determined for the stylolites in Bure-sur-Meuse borehole wells, from the crossover length scales, superposed with the current lithostatic stress and the maximum lithostatic stress reached during burial. After Rolland et al. (2012).

***Paleo stress magnitude in tectonic stylolites***: Following Ebner et al. (2009b) and Rolland et al. (2012), it is possible to determine tectonic stresses given formation depth, via the measured cross-over length scale $L^*$

$$L^* = \gamma E / \beta P \sigma_{SV} \qquad (25)$$

where $E$ is the Young's modulus of the rock, $\gamma$ is the surface tension, and the shear stress $\sigma_{SV}$ is the difference between the horizontal stress along y and the vertical stress along z,

$$\sigma_{SV} = \sigma_{yy} - \sigma_{zz} \qquad (26)$$

and when expressing,

$$P = (2\sigma_{zz} + \sigma_{yy})/3 \qquad (27)$$

in the absence of a net horizontal large-scale displacement, we can estimate from the crossover length scale $L^*$:

$$\sigma_{yy} = \frac{1}{2}\left(\sqrt{9\sigma_{zz}^2 + (12\gamma E/\beta L^*)} - \sigma_{zz}\right) \qquad (28)$$

This outcome allows us to determine the largest normal stress (Fig. 29). The largest horizontal stress determined, $\sigma_H = \sigma_{yy}$, between 10 and 17 MPa, exceeds the current vertical stress and the current horizontal ones at the depths (between 170 and 220 m) in the Dogger Formation. Rolland et al. (2014) concluded that these tectonic stylolites were formed during tectonic episodes at the Neogene and Paleogene periods, when horizontal stresses exceeded the vertical one at these depths. This methodology allows to determine the horizontal stress magnitude during such episodes.

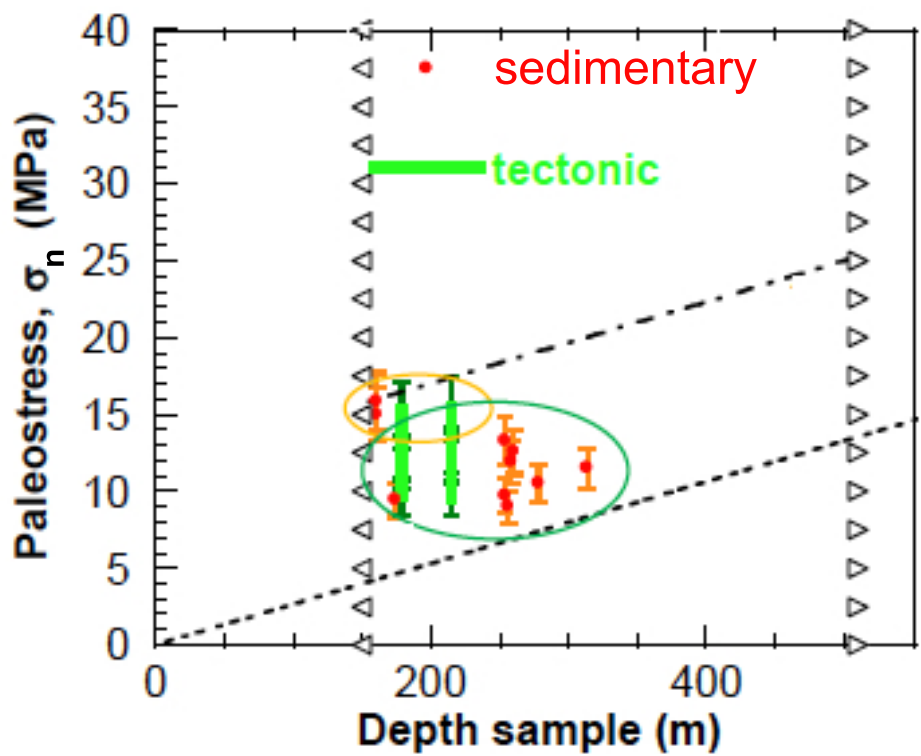

Figure 29: Normal stress determined across sedimentary and tectonic stylolites in the boreholes of the Bure-sur-Meuse Underground Rock Laboratory. Adapted after Rolland et al. (2012).

Beaudoin et al. (2016) compared stylolite stress analysis with paleopiezometry from calcite twinning in the Apennines, Italy. The authors show a strong correlation between tectonic stylolite and calcite twin stresses indicating that both methods complement each other.

*6.4 Evolution of stylolite transport properties – a challenge*

Stylolites are often believed to be an essential player in the compaction processes happening during diagenesis (e.g. Houseknecht, 1987, Spötl et al., 2000). Stylolitization processes are nonetheless not the only ones active during compaction, and are not always regarded as dominant at depth due to porosity decline (as modeled by the petroleum industry community – e.g., Lander and Walderhaug, 1999; Paxton et al., 2002). As stylolites develop, they modify the permeability by dissolving soluble minerals that then deposit in the near vicinity or further away, and sometimes by producing microcracks. Based on petrographic analyses and borehole logging data, it was suggested that sedimentary stylolites act as horizontal permeability barriers (e.g., Dunnington, 1967; Nelson, 1981; Burgess and Peter, 1985; Koepnick, 1987; Finkel and Wilkinson, 1990; Dutton and Willis, 1998; Alsharhan and Sadd, 2000), which may control reservoir permeability (e.g. Corwin, 1997). In contrast, several studies have observed enhanced porosity zones along the flanks and at the tip of the stylolite (Carozzi and von Bergen, 1987; Dawson, 1988; Raynaud and Carrio- Schaffhauser, 1992; Van Geet et al., 2000; Gingras et al., 2002; Harris, 2006), which led to the suggestion that stylolites actually enhance fluid flow (Carozzi and von Bergen, 1987; VanGeet et al., 2000). Since stylolites moreover merge together to form networks, and concentrate stress that leads to subsequent joint formation, subparallel or perpendicular to the stylolites, the connectivity of this dynamic networks of seals and drains may certainly play a key role on the transport properties of host sedimentary rocks, or for stressed regions, such as in the vicinity of faults. This behavior can be shown for stylolites in Zechstein carbonates in northern Germany where clay layers produce large teeth (Koehn et al., 2016). Ore minerals precipitate from fluids below flat parts of stylolite teeth where the stylolite acts as local seal and at the flanks of teeth where fluids breach the sealing stylolite.

Although various permeability investigations yield differing results regarding the sign of permeability change around stylolites, they agree that stylolites often affect both mechanics and flow. Heap et al. (2014) reported that stylolites in the Bure-sur-Meuse limestone drilling cores lead to a minor permeability increase in their vicinity, or no significant permeability change, depending on the sample tested. Baud et al. (2016) measured a porosity increase around these stylolites, presumably responsible for such permeability change. They also showed that this increase contributed to a measurable mechanical weakening of the rock, once stylolites exceed 5 mm thickness. In contrast, Corwin et al. (1997) found that stylolites in the Smackover Formation show more than 3 order of magnitude permeability reduction in the direction perpendicular to their plane. They attribute this decrease to a tight cement zone surrounding the stylolite, from a few millimeters to several centimeters thick, similar to Figure 17. Flow simulations reported by Corwin et al. (1997) suggest that although the cement zone around the stylolites is very thin, it controls large-scale flow and potential oil recovery. Lang et al. (2015) studied numerically the closing of rough fractures by pressuresolution, and showed how this leads to permeability decrease. Consequently, the question of when do stylolites induce permeability increase and when permeability decrease is not settled yet.

The permeability changes reported by Corwin et al. (1997) and Heap et al. (2014) are anisotropic, as are those predicted from recent numerical models (Koehn et al., 2016): permeability is either unmodified, in case of porosity increase adjacent to stylolite, or strongly decreased, in case of cement adjacent to stylolite, in the direction perpendicular to their average plane, and slightly increased or nearly unmodified in the direction along this plane. Heap et al. (2014) interpret permeability increase along the stylolitic plane to result from the network of fractures around the stylolites, and subparallel to them. They also suggested that the permeability perpendicularly to the stylolite plane might depend on the stylolite type and degree of maturation. They found hardly modified perpendicular permeability during experiments on thin stylolites, but for more mature stylolites, where low-permeability insoluble had accumulated into thick stylolites, a reduction of permeability perpendicular to the average stylolite plane could arise. Raynaud and Carrio-Schaffhauser (1992) identified a dissolution process zone around stylolites, increasing the surrounding porosity. On the other hand, cement derived from a dissolving stylolite will fill the forming process-zone pores and fractures emanating from the stylolites (e.g. Wangen, 1999; Walderhaug et al., 2006; Ben Itzhak et al., 2012; Angheluta et al., 2012). Koehn et al. (2016), in observations and numerical models, showed stylolites act either as seals and fluid pathways, depending on the insoluble material accumulating in the stylolite and the deformation of the sealing material at the teeth. The balance between opening of cracks and formation of process zone pores, increasing permeability, versus precipitation of cement, decreasing permeability, is probably a dynamic process. The transport properties of individual stylolites at different stages of maturation, with possible secondary fracture networks, is still an open research problem where experimental data for different types of stylolites would greatly benefit the understanding of their transport properties.

Similar to some stylolites, compaction bands (localized cataclasis) can act as permeability barriers in sedimentary formations (Baud et al., 2012; Deng et al., 2015). In sandstone, homogeneous grain size promotes the development of such localized features (Cheung et al., 2012).

**7. Conclusion and future challenges**

Our knowledge of stylolites has grown since Sorby's (1862) and Stockdale's (1922) works, with respect to detailed fractal morphology, evidence of cross-over length scales in these scaling laws, the better description of the physics related to their formation and evolution, and the ability to compare stylolite morphology with physics-driven numerical models. In addition, knowledge of permeability and microstructure related to stylolites and their surrounding rock has increased, and links have been found between these attributes and the stylolite evolution process. Indeed, a number of recent models based on physical principles reproduce stylolite morphology and stylolite networks, and help interpreting natural observations.

These models can explain how stylolites form through coupled mechanico-chemical processes. They provide the basis to understand what forces are responsible for the geometrical properties of stylolites at different spatial scales. In particular, they allowed identifying laws relating the cross-overs found in their morphology to their formation stress and the amount of shortening, across the stylolites. They open up the possibility for using stylolites as piezometers sensitive to the paleostress during their formation. Exactly which stress is measured (the stress during the early formation, the stress at maximum burial depth, or the stress during the most recent evolution of the stylolite) remains to be understood.

A question that is not solved is whether we have one, two or more different controlling mechanisms for stylolite evolution:

(i) A general dissolution process driven by differences in normal stress..

(ii) A local process driven by strain and surface energy

(iii) Clay, transport pathways, or grain-size controlled dissolution.

Numerous examples exist in nature and experiment where the main driving force is the difference in normal stresses. In addition, the explanation of the roughness by a competition between stabilizing terms (elastic or surface energy, depending on the scale) and destabilizing terms (dissolution disorder), seems convincing. Within this framework, a link remains to be made between dissolution surfaces with roughness (stylolites) and without roughness (cleavage, indenting.). They cannot have different driving forces. In addition, clays and even grain size are suggested to play a pivotal role in formation. Finally, all of these formulations assume reaction-controlled formation, yet some evidence suggests that at least sometimes formation is transport-controlled.

Several challenges can be identified for the future, notably; i) reproducing stylolites in the laboratory; ii) understanding (and using) the transport properties of stylolites; iii) understanding the control on the initiation of stylolites, i.e. the details of the driving force of the dissolution as function of stress and shape, to know under which conditions they can appear and interpret their presence; and iv) developing fully integrated models that will reproduce roughening, network and transport properties of stylolites.

It also remains to understand the problem of porosity evolution. We made the statement that if one assumes that the fluid is saturated with respect to the overall stress in the system, then the driving force for the normal stress term is only the stress difference along the interface and is as small as the Helmholtz free energy. Is there any porosity in this case? How would it evolve with time? Implementing models of this dynamic evolution of stylolite networks and the associated evolving rock transport properties is a current challenging and open problem.

*Acknowledgments*: We would like to thank Pascale Talour (CNRS and University of Grenoble Alpes), for her help to search for old scientific literature on stylolites. We thank K.L. Milliken, H. Fossen and W. Dunne for their valuable and careful reviews. We thank ANDRA for giving us access to valuable borehole samples from the Underground Research Laboratory in Bure-sur-Meuse. We acknowledge support from the EU FP7 ITN FlowTrans network, from the CNRS INSU ALEAS, from NEEDS MIPOR, and from the University of Strasbourg IDEX "Espoirs". This work was partly supported by the LIA France-Norway D-FFRACT, and the Research Council of Norway through its Centres of Excellence funding scheme, project number 262644.

**Notations**

| | |
|---|---|
| $A$ | thickness of material dissolved |
| $\alpha, \beta$ | multiplicative constants |
| $\Delta h(x, l)$ | height difference between two points at abcissas $x$ and $x + l$ |
| E | Young's modulus |
| $\varepsilon_p$ | plastic strain |
| $\eta$ | random term in model associated to each grain ("quenched noise") |
| $f_d$ | plastic energy stored in dislocations |
| $f_{el}$ | elastic strain energy |
| $f_s$ or $\psi_s$ | molar Helmholtz free energy |
| $f_{sc}$ | surface energy associated to surface charges and ions |
| $f_{se}$ | surface energy |
| $\gamma$ | surface free energy |
| $H$ | Hurst exponent |
| $h$ | height of a point, i.e. deviation from the average stylolite plane |
| $\tilde{h}_k$ | Fourier transform of height profile $h(x)$ |
| $k$ | wavenumber in Fourier transform |
| $\lambda$ | zoom factor in abcissa in an affine transformation |
| $L^*$ | cross-over length between small-scale and medium-scale scaling ranges |
| $l$ | separation between two points in abscissa |
| $\mu$ | chemical potential |
| P | fluid pressure, or pressure associated to far-field stress |
| $P(k)$ | Fourier power spectrum |
| $p_l(\Delta h)$ | distribution of height differences between points separated by $l$ in abscissa |
| $\sigma$ | stress tensor |
| $\sigma_n$ | stress component normal to an interface |
| $\sigma_s, \sigma_{SV}$ | far field shear stress |
| $\sigma_{gb}$ | normal stress acting on grain contacts |
| R | universal gas constant |
| $T$ | absolute temperature, in Kelvins |
| $V_s$ | molecular volume of solid mineral |
| $V_{sty}$, $v$ | dissolution speed |
| $w(l)$ | Characteristic height separation for two points $l$ away in abscissa |
| $w(x,l)$ | wavelet transform of profile |
| $X$ | cross-over length between medium-scale and large-scale scaling ranges |
| $x$ | Abcissa of a point in a stylolite profile |
| $\xi$ | correlation length |
| $z$ | dynamic exponent |
| $\Omega$ | Molar volume of a mineral |
| $\langle \rangle$ | spatial average over abcissas $x$ |

Hickman, S. H., Evans, B., 1995. Kinetics of Pressure Solution at Halite-Silica Interfaces and Intergranular Clay Films. Journal of Geophysical Research-Solid Earth, 100(B7), 13113-13132.

Hobbs, B. E., Means, W. D., and Williams, P. F., 1976. An outline of structural geology. Wiley and Sons.

Hopkins, T. C., 1897. Stylolites, American Journal of Science, 4, 142-144.

Houseknecht, D. W., 1987. Assessing the relative importance of compaction processes and cementation to reduction of porosity in sandstones. AAPG bulletin, 71(6), 633-642.

Iijima, A., 1979. Nature and origin of the Paleogene cherts in the Setogawa Terrain, Shizuoka, central Japan, J. Fac. Sci. Univ. Tokyo, Sect. 2, 20, 1-30.

Kaduri, M. 2013. Interconnected Stylolite Networks: field observations, characterization, and modeling, M.Sc. thesis, The Hebrew University of Jerusalem.

Kaplan, M. Y., 1976. Origin of stylolites. Docl. Acad. Sci USSR, Earth Sci. Sect., 221, 205-7.

Karcz, Z., Scholz, C. H., 2003. The fractal geometry of some stylolites from the Calcare Massiccio Formation, Italy. Journal of Structural Geology, 25(8), 1301-1316.

Karcz, Z; Aharonov, E; Ertas, D., Polizzotti, R., Scholz, C. H. 2006. Stability of a sodium chloride indenter contact undergoing pressure solution. Geology, 34(1), 61-63.

Karcz, Z.; Aharonov, E.; Ertas, D., Polizzotti, R., Scholz, C. H. 2008. Deformation by dissolution and plastic flow of a single crystal sodium chloride indenter: An experimental study under the confocal microscope. Journal of Geophysical Research-Solid Earth, 113, B4, B04205, doi 10.1029/2006JB004630.

Katsman, R., Aharonov, E., 2005. Modelling localized volume changes: Application to pressure solution and stylolites. Geochimica et Cosmochimica Acta, 69(10), A312-A312.

Katsman, R., Aharonov, E., Scher, H., 2006a. Localized compaction in rocks: Eshelby's inclusion and the Spring Network Model. Geophysical Research Letters 33(10), L10311, doi 10.1029/2005GL025628.

Katsman, R., Aharonov, E., Scher, H., 2006b. A numerical study on localized volume reduction in elastic media: Some insights on the mechanics of anticracks. Journal of Geophysical Research-Solid Earth 111(B3), B03204, doi 10.1029/2004JB003607.

Katsman, R., 2010. Extensional veins induced by self-similar dissolution at stylolites: analytical modeling. Earth and Planetary Science Letters, 299, 33-41.

Klöden, F., 1828. Beitrag zur Mineral. U. Geol. Kenntniss: der Mark Brandenburg, Bd., 1, 28.